# EXOPLANETS

**Space Science Series
University of Arizona Press**

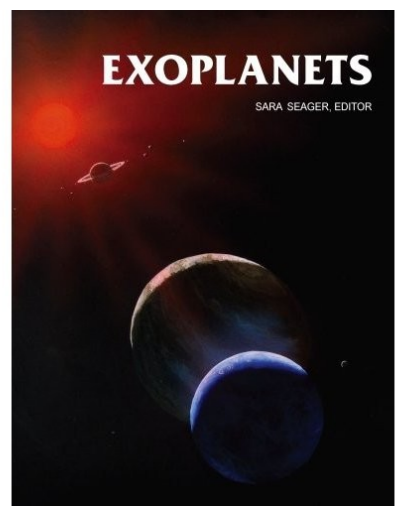

# Giant Planet Formation[†]

**Gennaro D'Angelo**
*NASA Ames Research Center and University of California, Santa Cruz*

**Richard H. Durisen**
*Indiana University*

**Jack J. Lissauer**
*NASA Ames Research Center*

Gas giant planets play a fundamental role in shaping the orbital architecture of planetary systems and in affecting the delivery of volatile materials to terrestrial planets in the habitable zones. Current theories of gas giant planet formation rely on either of two mechanisms: the core accretion model and the disk instability model. In this chapter, we describe the essential principles upon which these models are built and discuss the successes and limitations of each model in explaining observational data of giant planets orbiting the Sun and other stars.

## 1. INTRODUCTION

Jupiter and Saturn are composed predominantly of Hydrogen and Helium and are therefore referred to as *gas giants*, although most of these elements are not in gaseous form at the high pressures of these planets' interiors. The majority of extrasolar planets (or *exoplanets*) discovered so far have masses in excess of about one quarter of Jupiter's mass ($M_J$) and are known to be or suspected of being gas giants. Since Helium and molecular Hydrogen do not condense under conditions typically found in star forming regions and in protoplanetary disks, giant planets must have accumulated them as gasses. Therefore, giant planets must form prior to the dissipation of protoplanetary disks. Optically thick dust disks typically survive for only a few million years and protoplanetary disks lose all of their gaseous contents by an age of $\lesssim 10^7$ years (see the chapter by Roberge and Kamp), hence giant planets must form on this timescale or less.

Giant planets contain most of the mass and angular momentum of our planetary system, and thus they must have played a dominant role in influencing the orbital properties of smaller planets. Gas giants may also affect the timing and efficiency of the delivery of volatile materials (*Chambers and Wetherill*, 2001), such as water and carbon compounds, to the habitable zones of planetary systems, where liquid water can exist on a (rocky) planet's surface. In the solar system, Jupiter is also believed to have reduced the impact rate of minor bodies (such as comets) on the Earth (*Horner et al.*, 2010). Therefore, understanding giant planet formation is essential for the formulation of theories describing the origins and evolution of terrestrial planets capable of sustaining life in the form of complex organisms.

Observations of exoplanets have vastly expanded our database by increasing the number of known planets by over one and a half orders of magnitude. The distribution of observed exoplanets is highly biased towards those objects that are most easily detectable using the Doppler radial velocity and transit photometry techniques (see chapters by Lovis and Fischer and by *Winn*), which have been by far the most effective methods of discovering planetary-mass object orbiting other stars. Although these extrasolar planetary systems are generally different from the solar system, it is not yet known whether our planetary system is the norm, quite atypical, or somewhere in between. Nonetheless, some unbiased statistical information can be distilled from available exoplanet data (see the chapter by Cumming).

The mass distribution function of young compact objects in star forming regions (e.g., *Zapatero Osorio et al.*, 2000) extends down through the brown dwarf mass range to below the deuterium burning limit (12–14 $M_J$). This observed continuity implies that most isolated brown dwarfs and isolated high planetary-mass objects form via the same collapse process as do stars.

However, star-like direct quasispherical collapse is not considered a viable mechanism for the formation of Jupiter-mass planets, because of both theoretical arguments and observational evidence. The *brown dwarf desert*, a profound dip over the range from $\sim 15\, M_J$ to $\sim 60\, M_J$ in the mass distribution function of companions orbiting within several AU of Sun-like stars, suggests that the vast majority of gas giants form via a mechanism different from that of stars.

---

[†] In EXOPLANETS, edited by S. Seager, to be published in the Fall of 2010 in the Space Science Series of the University of Arizona Press (Tucson, AZ).

A theory based on a unified formation scenario for rocky planets, ice giants much as Uranus and Neptune, and gas giants is the *core accretion model*, in which the initial phases of a giant planet's growth resemble those of a terrestrial planet's. The only alternative formation scenario receiving significant attention is the *disk instability model*, in which a giant planet forms directly from the contraction of a clump of gas produced via a gravitational instability in the protoplanetary disk.

In this chapter, we introduce the basic physical concepts of gas giant planet formation according to core accretion models (Section 2) and disk instability models (Section 3). In Section 4, we present theoretical arguments suggesting that most of the giant planets known to date formed via core accretion and pose some of the outstanding questions that still need an answer. Future prospects that may settle some of the lingering issues are discussed in Section 5.

## 2. CORE ACCRETION MODELS

The initial phases of a giant planet's growth by core nucleated accretion proceed through an accumulation process of solid material, in the same fashion as terrestrial planets form. Dust and small solid grains, which are entrained in the predominant gas component of a protoplanetary disk, coagulate into larger particles. Centimeter-sized particles tend to settle towards the disk midplane, aggregating and eventually forming kilometer-sized agglomerates, referred to as *planetesimals*.

Planetesimals grow larger via pairwise collisions, leading to the formation of a *planetary embryo*. An embryo may have the ability to grow at a rate that increases as its mass increases, eventually consuming nearly all planetesimals in the neighborhood, while rapidly gaining mass and becoming a *planetary core* or *protoplanet*. Once a planetary core grows large enough and the escape velocity from its surface exceeds the thermal speed of the surrounding gas, a tenuous envelope of gas begins to accumulate around the solid core.

For most of the following planet's growth history, thermal pressure effects within the envelope regulate the accretion of gas. The ability of the envelope to radiate away the gravitational energy released by incoming planetesimals and by contraction limits the amount of gas that can be accreted by the planet. A slow contraction phase ensues in which the accretion rates of both solid material and gas are small. However, as the protoplanet grows and its total mass exceeds the value beyond which the pressure gradient (in the envelope) can no longer balance the gravitational force, the envelope undergoes a phase of rapid contraction, which allows more gas to be accreted. The augmented mass of the envelope triggers further contraction and gas can be thereby accreted at an ever increasing rate.

During this epoch, known as runaway gas accretion phase, the gas accretion rate is regulated by the ability of the surrounding disk to supply gas to the planet's vicinity. In this stage, once the planet's mass exceeds a few tenths of Jupiter's mass (for typical temperature and viscosity conditions in a protoplanetary disk around a solar-mass star), the gas accretion rate decreases as the planet mass increases due to tidal interactions between the planet and the disk. The giant planet continues to grow at an ever decreasing rate until there is no gas available within the planet's gravitational reach.

### 2.1. From Dust to Planetesimals

The formation of a heavy element core is an essential part in any core accretion model. Hence, for completeness, we present a summary of the basic elements involved in this process. An in-depth discussion of these concepts can be found in the chapter by Chambers.

The core formation starts at some distance from the star, most likely beyond the "snow line" where disk temperatures allow for condensation of water ice and solid material. This process begins from (sub-)micron-sized dust particles, which may have originated from the interstellar medium and/or condensed within the disk's gas. Such small solid particles are well-coupled to the gas, on account of their large surface area to mass ratio, and therefore move with it.

As particles collide and stick together, they can grow larger and start to de-couple from the gas and interact with it because of differential rotation between solids and gas. The relative velocity stems from the fact that the gas rotates about the star slower than do solids due to the radial pressure gradient that partially counteracts the gravitational attraction of the gas towards the star. The resulting interaction can be described in terms of friction via gas drag on the particle, producing an acceleration proportional to and in the opposite direction of the relative velocity between the particle and the gas. This friction generally removes orbital angular momentum from solid particles, causing them to drift towards the star.

Along with the radial drift, small solids also experience a friction in the direction perpendicular to the disk midplane. Assuming that the vertical motion of the gas is negligible (compared to that of solids), the vertical motion of a solid particle is obtained by applying the second law of dynamics

$$\frac{d^2 z}{dt^2} + \frac{1}{\tau_f}\frac{dz}{dt} + z\,\Omega^2 = 0\,, \quad (1)$$

in which $z$ represents the distance over the disk midplane, $\tau_f$ is the *friction timescale* (also referred to as "stopping time"), and $\Omega$ is the particle's angular velocity about the star. The second term on the lefthand side of equation (1) is the opposite of the frictional force per unit mass exerted by the gas and the third term is the opposite of the vertical gravitational acceleration imposed by the star. In the limit $\tau_f \to 0$, dust is perfectly coupled to gas and $dz/dt$ is equal to the vertical gas velocity, which is zero by assumption. Hence $z$ is a constant and no sedimentation would occur. Notice that for the gaseous part of the disk, equation (1) is replaced by an equation for hydrostatic equilibrium in the vertical direction, with the vertical component of the gas pressure gradient balancing the vertical gravitational force (see equation (3)). In the limit $\tau_f \to \infty$, equation (1) assumes the form of a harmonic oscillator, hence particles oscillate about the disk midplane with a period equal to $2\pi/\Omega$.

For solid particles whose size is shorter than the mean free path of gas molecules and whose velocity relative to the gas is slower than the gas sound speed, $c$, the friction time is (*Epstein*, 1924)

$$\tau_f = \left(\frac{\rho_{\rm sp}}{\rho}\right)\left(\frac{R_{\rm sp}}{c}\right), \qquad (2)$$

where $\rho_{\rm sp}$ and $R_{\rm sp}$ are the solid particle's volume density and radius, respectively, while $\rho$ is the gas density. Indicating with $H$ the vertical scale-height of the gaseous component of the disk and with $a$ the orbital distance to the star, hydrostatic equilibrium in the vertical direction requires that

$$\frac{1}{\rho}\frac{\partial p}{\partial z} + \left(\frac{GM_\star}{a^2}\right)\left(\frac{z}{a}\right) = 0, \qquad (3)$$

in which $p = p(a,z)$ is the gas pressure and $M_\star$ is the mass of the star. Writing the pressure as $p = c^2\rho$, approximating $c$ as being independent of height, and integrating over the disk thickness one finds that $c^2 \approx H^2\Omega^2$, where $\Omega = \sqrt{GM_\star/a^3}$ is the Keplerian angular velocity of the gas. (A more accurate determination of the disk's rotation rate can be derived from imposing hydrostatic equilibrium in the radial direction, which results in $\Omega^2 \simeq (GM_\star/a^3)[1-(H/a)^2]$). Thus, equation (2) gives

$$\Omega\tau_f = \left(\frac{a}{H}\right)\left(\frac{\rho_{\rm sp}}{\rho}\right)\left(\frac{R_{\rm sp}}{a}\right). \qquad (4)$$

Assuming typical values at $a \approx 5$ AU for $\rho$ and $H/a$ around a solar-mass star of $10^{-10}$ g cm$^{-3}$ and 0.05, respectively, and expressing both the density and radius of the particle in cgs units, $\Omega\tau_f \sim 10^{-3}\rho_{\rm sp}R_{\rm sp}$. This estimate is applicable up to values of $R_{\rm sp}$ less than tens of centimeters, i.e., on the order of the mean free path of gas molecules ($\propto 1/\rho$) under the adopted gas conditions.

Equation (1) has the form a damped harmonic oscillator and can be integrated once initial conditions are provided for position and velocity. Setting $z = z_0$ and $\dot{z} = dz/dt = 0$ at time zero and taking into account the inequality $\Omega\tau_f \ll 1$ found above, the solution can be approximated as

$$z \simeq z_0\left(e^{-t\Omega^2\tau_f} - \Omega^2\tau_f^2 e^{-t/\tau_f}\right). \qquad (5)$$

The second term in the above solution is a fast transient that rapidly decays to zero, hence we can further approximate the solution as $z \approx z_0 e^{-t\Omega^2\tau_f}$. Therefore, in order for the altitude above the midplane to reduce by more than 99% of its initial value, $\Omega t > 5/(\Omega\tau_f) \sim 5 \times 10^3/(\rho_{\rm sp}R_{\rm sp})$, in which again $\rho_{\rm sp}$ and $R_{\rm sp}$ are expressed in cgs units. Micron-sized particles would take millions of orbital periods ($\tau_{\rm rot} = 2\pi/\Omega$) to settle, which suggests that they first need to aggregate and grow into larger particles.

Centimeter-sized icy/rocky aggregates would require only on the order of thousands of orbits to approach the disk's midplane.

A thin layer of solid material may thus accumulate at the disk's midplane in a relatively short amount of time compared to disk lifetimes. If this layer is sufficiently dense, clumps may form through gravitational instabilities within the layer (e.g., *Goldreich and Ward*, 1973). Such process may produce kilometer-sized bodies, known as planetesimals.

The assumption that the gas vertical velocity is small compared to the particle vertical velocity is, however, only valid in absence of sustained turbulent motions. In general, protoplanetary disks are expected to be somewhat turbulent and moderate amounts of turbulence could affect the settling timescales of small grains ($R_{\rm sp} \lesssim 1$ cm). Ignoring the second term in equation (5), particle speeds are $\dot{z} \approx -z\Omega^2\tau_f$. Assuming a turbulent kinematic viscosity of the form $\nu = \alpha H c$ (*Shakura and Syunyaev*, 1973) and indicating with $\lambda_{\rm turb}$ the vertical mixing length (the typical size of eddies), vertical gas speeds due to turbulence can be estimated as $|\dot{z}_{\rm turb}| \sim \nu/\lambda_{\rm turb}$ or $|\dot{z}_{\rm turb}| \sim \alpha H c/\lambda_{\rm turb} = \alpha H^2\Omega/\lambda_{\rm turb}$, and thus

$$\left|\frac{\dot{z}}{\dot{z}_{\rm turb}}\right| \sim \left(\frac{z\lambda_{\rm turb}}{H^2}\right)\left(\frac{\Omega\tau_f}{\alpha}\right). \qquad (6)$$

Since $z\lambda_{\rm turb}/H^2 \leq 1$ (eddies cannot be larger than the disk's thickness), we have that $|\dot{z}/\dot{z}_{\rm turb}| < \Omega\tau_f/\alpha$. Notice that in order for turbulent motions to be subsonic (i.e., $|\dot{z}_{\rm turb}| \leq c$), the turbulent kinematic viscosity $\nu \leq c\lambda_{\rm turb} \leq cH$ and thus $\alpha \leq 1$. Under the disk's conditions adopted above, the ratio of the particle's settling velocity to the gas vertical turbulent speed is then $|\dot{z}/\dot{z}_{\rm turb}| < 10^{-3}\rho_{\rm sp}R_{\rm sp}/\alpha$ ($\rho_{\rm sp}$ and $R_{\rm sp}$ are in cgs units).

Therefore, if the turbulence parameter[1] $\alpha \gtrsim 10^{-3}$, the settling time of centimeter-sized particles may be affected since vertical turbulent mixing could influence their vertical transport. Studies of grain settling in the presence of turbulent motions indicate that particles tend to concentrate in stagnant regions of the flow and that concentrations are size-dependent, which may lead to the accumulation of sub-centimeter-sized particles (*Cuzzi et al.*, 2001). Previous growth of these agglomerates would still rely on sticking collisions of smaller particles.

The growth from centimeter-sized to kilometer-sized bodies is still a poorly understood process and an active field of both theoretical and experimental research. Nonetheless, there is observational evidence that it does occur in nature: dust particles are observed in debris disks around other stars and small bodies of tens to hundreds of kilometers in size are observed in the Kuiper belt around the Sun.

---

[1] Based on observed gas accretion rates and other properties of young stellar objects with disks, typical values of $\alpha$ are in the range from $\sim 10^{-4}$ to $\sim 0.1$ (e.g., *Hueso and Guillot*, 2005; *Isella et al.*, 2009).

## 2.2. From Planetesimals to Planetary Cores

We shall now assume that the solid component of the protoplanetary disk has had time to agglomerate into planetesimals, rocky/icy bodies of a kilometer (or larger) in size. These objects are massive enough ($\gtrsim 10^{15}$–$10^{16}$ g $\sim 10^{-12}\,M_\oplus$) that they may gravitationally interact with their neighbors and perturb their velocities. As a result of these interactions, planetesimals become prone to collisions.

Although the outcome of a collision between two planetesimals depends upon their relative velocity, we assume that collisions among such objects result in mergers rather than fragmentation. Under this assumption, the growth rate of a planetesimal of mass $M_s$ can be written as (*Safronov*, 1969)

$$\frac{dM_s}{dt} = \pi R^2\, v_\mathrm{rel}\, \rho_s\, F_g, \quad (7)$$

where $R$ is the planetesimal radius, $v_\mathrm{rel}$ is the relative velocity between the two impacting bodies, $\rho_s$ is the volume density of the solid component of the disk. The product $v_\mathrm{rel}\rho_s$ represents a mass flux, that is the amount of solid material sweeping across the target planetesimal per unit time and unit area. The quantity $F_g = (R_\mathrm{eff}/R)^2$ is a gravitational enhancement factor, which is the ratio of the effective cross-section ($\pi R_\mathrm{eff}^2$) of the accreting planetesimal to its geometrical cross-section ($\pi R^2$). This factor accounts for the ability of the growing body to bend towards itself the trajectories of other, sufficiently close, planetesimals.

It is often useful to express the accretion rate $dM_s/dt$ in terms of the surface density of the solid material $\Sigma_s = \int \rho_s dz \sim H_s \rho_s$, with $H_s$ being the vertical thickness of the planetesimal disk. In order to do so, one can assume that the gravitational force exerted by the star is the major component of the force acting on planetesimals in the vertical direction of the disk (as in equations (1) and (3)) and that relative velocities between planetesimals are isotropic. Thus, $H_s$ is of order $v_\mathrm{rel}/\Omega$, where $\Omega$ is the Keplerian angular velocity of the growing planetesimal. Equation (7) can then be cast in the form $dM_s/dt = \pi R^2 \Omega \Sigma_s F_g$. Since the angular velocity along a Keplerian orbit is $\Omega \propto a^{-3/2}$, the accretion rate of planetesimals is slower farther from the star (neglecting variations of $\Sigma_s F_g$ with distance). This would imply that $dM_s/dt$ at the current location of Uranus (19.2 AU) was about 7 times as small as it was at the current location of Jupiter (5.2 AU).

If we neglect the collective gravitational action of the other planetesimals and that of the star during an encounter, two interacting bodies can be described in the framework of a two-body problem. Assuming that the target planetesimal has already grown somewhat larger than the neighboring bodies, hence becoming a planetary embryo, the impacting body can be thought of as a projectile. We can therefore use the approximation that the embryo sits on the center of mass of the two-body system. In the rest frame of the embryo, conservation of the specific angular momentum (i.e., angular momentum per unit mass) reads $R_\mathrm{eff}\, v_\mathrm{rel} \simeq R\, v_\mathrm{ta}$, where $v_\mathrm{ta}$ is the relative velocity for a tangential approach (when the projectile grazes the embryo). Conservation of specific energy during the collision requires that $v_\mathrm{rel}^2 \approx v_\mathrm{ta}^2 - 2GM_s/R$ and therefore the gravitational enhancement factor, i.e., the ratio $(R_\mathrm{eff}/R)^2$, can be cast in the form

$$F_g = 1 + \frac{v_\mathrm{esc}^2}{v_\mathrm{rel}^2}, \quad (8)$$

in which $v_\mathrm{esc} = \sqrt{2GM_s/R}$ is the escape velocity from the surface of the target planetesimal. Notice that, if the planetesimal radius is not negligible compared to the embryo radius, the radius $R$ in equation (7) and in $v_\mathrm{esc}$ should be replaced by the sum of the two radii.

If relative velocities are high and $v_\mathrm{rel} \gg v_\mathrm{esc}$, then $F_g \approx 1$ (i.e., $R_\mathrm{eff} \approx R$) and equation (7) yields $dM_s/dt \propto R^2 \propto M_s^{2/3}$. If relative velocities are low and $v_\mathrm{rel} \ll v_\mathrm{esc}$, then $F_g \approx v_\mathrm{esc}^2/v_\mathrm{rel}^2 \propto R^2$ and $dM_s/dt \propto R^4 \propto M_s^{4/3}$. If the escape velocity is very much larger than the relative velocity, then three-body effects (star, planetary embryo, and planetesimal), neglected in deriving equation (8), must be taken into account to compute $F_g$ (*Greenzweig and Lissauer*, 1992).

In the high relative velocity regime, the growth timescale $\tau_s = M_s(dM_s/dt)^{-1}$ (i.e., the time it takes for the mass of the embryo to increase by a factor $e \approx 2.7$), is proportional to $M_s^{1/3}$ and therefore the growth rate, $1/\tau_s$, of an embryo reduces as it grows larger. This implies an *orderly growth* of large planetesimals, which tend to attain similar masses. From equation (7) cast in terms of solids' surface density, one obtains that the timescale $\tau_\mathrm{og}$ necessary to build a large embryo, or planetary core, of mass $M$ can be estimated as

$$\Omega \tau_\mathrm{og} \sim \frac{1}{\pi}\left(\frac{4\pi}{3}\right)^{2/3} \frac{(\rho_\mathrm{sp}^2 M)^{1/3}}{\Sigma_s}. \quad (9)$$

To assemble a body of mass $M \sim 1.6 \times 10^{-4}\,M_\oplus$ (about as massive as Ceres, the largest object in the asteroid belt) within a solids' disk of density $10\,\mathrm{g\,cm}^{-2}$, orderly growth would require a few $10^6$ orbital periods, which already represents a fairly long timescale compared to lifetime of protoplanetary disks around solar-mass stars ($\lesssim 10^7$ years).

In the low relative velocity regime, the growth rate of an embryo is $1/\tau_s \propto M_s^{1/3}$, and thus the larger the planetary embryo the faster it grows, a process known as *runaway growth*. During runaway growth, the largest embryo grows faster than any other embryo within its accretion region. Although a two-body approximation (equation (8)) yields an unlimited gravitational enhancement factor as the ratio $v_\mathrm{rel}/v_\mathrm{esc} \to 0$, gravitational scattering due to three-body effects set a limit to $F_g$, which cannot exceed values much beyond several thousands (*Lissauer*, 1993).

During the assembly of a planetary core through the growth of an embryo, relative velocities among planetesimals play a crucial role in determining the accretion rates, as indicated by equations (7) and (8). The velocity distribution of a swarm of planetesimals is affected by a number of physical processes, such as elastic and inelastic collisions, gravitational scattering, and frictional drag by the gas, the results of which can be highly stochastic.

If the orbit radial excursion of planetesimals, relative to the orbit of an embryo, is on the order of the embryo's Hill radius[2]

$$R_{\mathrm{H}} = a \left( \frac{M_s}{3M_\star} \right)^{1/3}, \quad (10)$$

or smaller, then the accretion is said to be shear-dominated because growth is dictated by Keplerian shear in the disk, rather than by planetesimals' random velocities. If the embryo's orbit is nearly circular, this situation requires that orbital eccentricities and inclinations of planetesimals should be $\sim R_{\mathrm{H}}/a$, or smaller. The relative velocity between an embryo and a planetesimal traveling on a circular Keplerian orbit with radii $a$ and $a + \Delta a$ ($\Delta a/a \ll 1$), respectively, is $v_{\mathrm{rel}} \simeq a|\Omega(a+\Delta a) - \Omega(a)| \approx a\Omega(a)|[1-3\Delta a/(2a)] - 1|$, hence $|v_{\mathrm{rel}}| \approx 3\Omega \Delta a/2$. If we approximate $\Delta a$ as the half-width of the region within which the gravity field of the embryo dominates over that of nearby embryos, the accretion rate is $dM_s/dt \propto M_s^{2/3} \Omega \Delta a \, \rho_s F_g$. We will see in Section 2.3 that $\Delta a$ is generally proportional to the embryo's Hill radius, $R_{\mathrm{H}}$. Since $|v_{\mathrm{rel}}| \sim \Omega R_{\mathrm{H}}$ and $\rho_s \sim \Sigma_s/H_s \sim \Sigma_s/R_{\mathrm{H}}$, the planetesimal accretion rate becomes $dM_s/dt \propto M_s^{2/3} \Omega \Sigma_s F_g$.

This result is formally the same as the accretion rate in the orderly regime (in which $v_{\mathrm{esc}}^2/v_{\mathrm{rel}}^2 \ll 1$ and $F_g \approx 1$), except that now $v_{\mathrm{esc}}^2/v_{\mathrm{rel}}^2 \sim 6R_{\mathrm{H}}/R \gg 1$, and therefore $F_g \gg 1$. Accordingly, the growth timescale (in units of $1/\Omega$) is given by the righthand side of equation (9) divided by the gravitational enhancement factor, $F_g$. The growth rate of an embryo is much larger than it is during the orderly growth phase (since $F_g \gg 1$), but it reduces as the embryo mass grows larger (see equation (9)). This phase of growth, often referred to as *oligarchic*, may lead to the formation of massive embryos at regular intervals in semi-major axis. Notice that, neglecting variations of $\Sigma_s F_g$ with orbital distance, as in the orderly growth regime $dM_s/dt \propto M_s^{1/2} a^{-3/2}$. This implies that the accretion rate of planetesimals reduces as the stellar mass decreases or as the distance from the star increases.

### 2.3. Isolation Mass of Planetary Cores

The rapid (runaway/oligarchic) growth of a planetary embryo continues until its neighborhood, or *feeding zone*, has been substantially cleared of planetesimals (*Lissauer*, 1987). The feeding zone represents the domain within which the dominant embryo is able to significantly deflect the paths of other planetesimals towards itself. The radial extent of this region is several Hill radii, $R_{\mathrm{H}}$, which can be understood by recalling the definition of $R_{\mathrm{H}}$. This is the distance of the equilibrium point from mass $M_s$ (the embryo), on the line connecting (and in between) masses $M_\star$ and $M_s$, in a reference frame rotating at angular velocity $\Omega$. The force balance for a planetesimal of mass much smaller than both $M_\star$ and $M_s$ requires that $M_s R_{\mathrm{H}}^{-2} - M_\star(a - R_{\mathrm{H}})^{-2} + \Omega^2(a - R_{\mathrm{H}}) = 0$. Hence, to leading order in $R_{\mathrm{H}}/a$, $M_s R_{\mathrm{H}}^{-2} - M_\star a^{-2}(1 + 2R_{\mathrm{H}}/a) + M_\star a^{-2}(1 - R_{\mathrm{H}}/a) = 0$, whose solution is the righthand side of equation (10).

Once the feeding zone has been severely depleted, the planetary embryo becomes nearly isolated. Numerical N-body simulations suggest that isolation of an embryo occurs once a region of width $\Delta a \sim b R_{\mathrm{H}}$, where $b \approx 4$, on each side of the embryo's orbit becomes nearly emptied of planetesimals (*Kokubo and Ida*, 2000). The *isolation mass*, $M_{\mathrm{iso}}$, of a planetary embryo can therefore be calculated as $M_{\mathrm{iso}} \sim 4\pi \, a \, \Delta a \, \Sigma_s \sim 4\pi \, a \, b \, R_{\mathrm{H}} \, \Sigma_s$. Notice that here $\Sigma_s$ refers to the *initial* value of the solids' surface density. By using equation (10), the isolation mass can be written as

$$M_{\mathrm{iso}} \sim \sqrt{\frac{(4\pi \, a^2 \, b \, \Sigma_s)^3}{3 M_\star}}. \quad (11)$$

According to equation (11), a planetary core of mass $\sim 11\,M_\oplus$ (if $b \approx 4$) would become nearly isolated at 5.2 AU from a solar-mass star if the local surface density of solids were equal to $10\,\mathrm{g\,cm^{-2}}$. This value for $\Sigma_s$ is a couple of times as large as that predicted for a minimum mass proto-solar nebula[3] (MMSN). Additionally, the isolation mass increases with distance from the star, although this increase can be somewhat compensated for by a reduced surface density of solids, which is expected to decrease as $a$ increases.

An order-of-magnitude estimate of the timescale, $\tau_{\mathrm{iso}}$, required to reach isolation can be obtained by taking the ratio of equation (11) to the planetesimal accretion rate (expressed in terms of $\Sigma_s$, see Section 2.2), which yields

$$\Omega \tau_{\mathrm{iso}} \sim \frac{\mathcal{C}_{\mathrm{iso}}}{F_g} \sqrt{\frac{M_\star}{a^2 \Sigma_s}} \left( \frac{a^3 \rho_{\mathrm{sp}}}{M_\star} \right)^{2/3}, \quad (12)$$

where $\mathcal{C}_{\mathrm{iso}} = (2\pi)^{2/3} (2/\sqrt{3})^{5/3} \sqrt{b/\pi} \simeq 2.44\sqrt{b}$. Notice that for given values of $F_g$ and $\Sigma_s$, this timescale increases with increasing distance from the star and decreases with increasing stellar mass: $\tau_{\mathrm{iso}} \propto a^{5/2} M_\star^{-2/3}$. At the current location of Jupiter around a solar-mass star, $\Omega \tau_{\mathrm{iso}} \sim 3 \times 10^8/F_g$, if $\Sigma_s$ is $10\,\mathrm{g\,cm^{-2}}$. Assuming a situation in which the gravitational enhancement factor is of order 1000 over the entire accretion epoch, $\tau_{\mathrm{iso}}$ is several tens of thousands of orbital periods or several $10^5$ years.

---

[2]The Hill radius, $R_{\mathrm{H}}$, represents the distance of the Lagrange point $L_1$ from the secondary in the circular restricted three-body problem (e.g., *Murray and Dermott*, 2000). It provides a rough measure of the distance from the secondary beyond which the gravitational attraction of the primary and centrifugal effects prevail over the gravity of the secondary. Note however that this region, which identifies the Roche lobe, is not a sphere and its volume is about a third of a sphere of radius $R_{\mathrm{H}}$ (e.g., *Eggleton*, 1983).

[3]This is defined as the amount of heavy elements (heavier than Helium) observed in the planets and minor bodies of the solar system (mostly contained in giant planets) augmented by an amount of gas such to render the proto-solar nebula composition equal to that of the young Sun. Such definition constrains the total mass, $\sim 0.02\,M_\odot$, and, to a lesser extent, the surface density distribution of the minimum mass solar nebula (see, e.g., *Davis*, 2005).

In deriving equation (12) we assumed that the surface density of solids, $\Sigma_s$, is comparable to its initial value and the geometrical cross-section of the planet only depends on the core radius, $R$. Yet, the surface density of solids drops if accreted planetesimals are not replaced by others arriving from outside of the core's feeding zone, which would operate towards increasing $\tau_{\rm iso}$. Moreover, during its growth toward isolation, a planetary core also accretes gas from the disk, although at a much smaller rate than it accretes solids. Once the atmospheric envelope becomes massive enough to dissipate the kinetic energy of incoming planetesimals via drag friction, the geometrical cross-section of the protoplanet in equation (7) becomes substantially larger than $\pi R^2$. This effect would operate towards reducing $\tau_{\rm iso}$.

A planetary core does not necessarily stop growing once attaining the mass given by equation (11). Perturbations among planetesimals and other embryos can supply additional solid material to the core's feeding zone via scattering. Furthermore, a planet in excess of a Mars mass can exert gravitational torques on the surrounding gas. The disk responds by exerting the same amount of torques onto the planet, which modify the planet's orbital angular momentum, forcing it to radially migrate within the disk. As a result of this radial displacement, a planet may reach disk regions that still contain planetesimals. However, as the planet's mass grows, so does its ability to scatter planetesimals away from the orbital path or to trap them into mean-motion resonances (so that the ratio of a planetesimal's orbital period to the planet's orbital period is a rational number).

The order-of-magnitude estimates given above for the isolation mass and the timescale to reach it neglect many aspects of the physical processes involved in the growth of a planetary core. Nonetheless, detailed calculations of giant planet formation by core accretion and gas capture indicate that those estimates are valid under appropriate conditions. Calculations that start from a planetary embryo of $M_s \sim 0.1\,M_\oplus$ (about equal to the mass of Mars) orbiting a solar-mass star at $5.2\,{\rm AU}$ and undergoing rapid growth within a planetesimal disk of initial surface density $\Sigma_s \approx 10\,{\rm g\,cm^{-2}}$, show that a planetary core becomes nearly isolated within less than half a million years when $M_{\rm iso} \approx 11\,M_\oplus$ (Pollack et al., 1996), as also indicated by the solid line in left panel of Figure 1 (see Figure's caption for further details).

## 2.4. Growth of Thermally-Regulated Envelopes

Gas can accrete onto a planetary embryo when the thermal energy is smaller than the gravitational energy binding the gas to the embryo. This condition is satisfied when the escape velocity from the surface of the embryo, $v_{\rm esc}$ (see Section 2.2), exceeds the local thermal speed of the disk's gas ($\sqrt{8/\pi}c$), which occurs when $M_s \gtrsim 4(H/a)^2 M_\star R/(\pi a)$, or

$$M_s \gtrsim \sqrt{\frac{M_\star^3}{a^3 \rho_{\rm sp}}} \left(\frac{H}{a}\right)^3. \tag{13}$$

At distances of several AU from a solar-mass star, relatively small bodies ($M_s \sim 0.01\,M_\oplus$) can retain an atmosphere. In these early phases, the atmospheric gas is optically thin and thermal energy released by impacting planetesimals can be readily radiated away, allowing for contraction of the envelope and for additional gas to be accreted.

Prior to achieving isolation from the planetesimal disk, the accretion rate of solids is much larger than that of gas. The growing core mass forces the envelope to contract and more gas can be collected from the surrounding gaseous disk. However, as the envelope grows more massive, it becomes increasingly optically thick to its own radiation, which therefore cannot escape to outer space as easily as in earlier phases. As a consequence, envelope's temperature and density exceed those of the local disk's gas.

The pressure gradient that builds up in the envelope opposes gravitational contraction, preventing accretion of large amounts of gas. On the one hand, ongoing accretion of solid material and growth of the core help envelope contraction, on the other the gravitational energy released in the envelope by gas compression and that supplied by accreted planetesimals help maintain a relatively large pressure gradient. Therefore, contraction becomes *self-regulated*. Once the planet achieves (near) isolation from planetesimals, the accretion rate of solids becomes small and accretion of gas continues to the extent allowed by envelope compression.

The envelope enters a stage of quasistatic contraction that can be characterized by long evolution timescales. If the accretion of solids and gas was negligibly small, the timescale over which the envelope contracts would be related to the ratio of the gravitational energy released by contraction, $|E_{\rm grav}|$, to the envelope's luminosity, $L$. This ratio defines the Kelvin-Helmholtz timescale $\tau_{\rm KH} = |E_{\rm grav}|/L$. Indicating the protoplanet's core and envelope masses with $M_c$ and $M_e$, respectively, $|E_{\rm grav}| \sim G M_c M_e / \bar{R}$, $\bar{R}$ being the average radius comprising most the protoplanet's mass, and thus the contraction timescale is

$$\tau_{\rm KH} \sim \frac{G\,M_c M_e}{\bar{R}\,L}. \tag{14}$$

Using values from Figures 1 and 2 around the middle of the planetesimal isolation phase, $\tau_{\rm KH}$ would be of order $10^5$ years. However, for the case illustrated in the Figures, most of the luminosity produced during this slow growth phase is due to gravitational energy generated by accretion rather than by contraction of the envelope.

The length of this epoch depends on several factors, but principally on the opacity of the envelope and on the solids' accretion rate. The more opaque the gas is to outgoing radiation, the less able the envelope is to dissipate the energy provided by gas compression and/or by continued planetesimal accretion. Moreover, accreted solids can be consumed by ablation during their atmospheric entry, releasing dust and increasing the envelope's opacity. Therefore, a reduced accretion of solids shortens this epoch. As shown in the left panel of Figure 1, the slow contraction phase, from the time the core mass reaches near-isolation from surrounding plan-etesimals to the time the rapid gas accretion phase begins, lasts about 2 Myr. The assumption made for this particular model is that the grain opacity in the protoplanet's envelope is 2% that of the interstellar medium and that planetesimals continue to accrete throughout the slow contraction phase. All else being equal, this time-span increases to 6 Myr when the full interstellar opacity is adopted and decreases to $\approx 1\,\mathrm{Myr}$ if planetesimal accretion ceases entirely once the isolation mass is achieved (*Hubickyj et al.*, 2005).

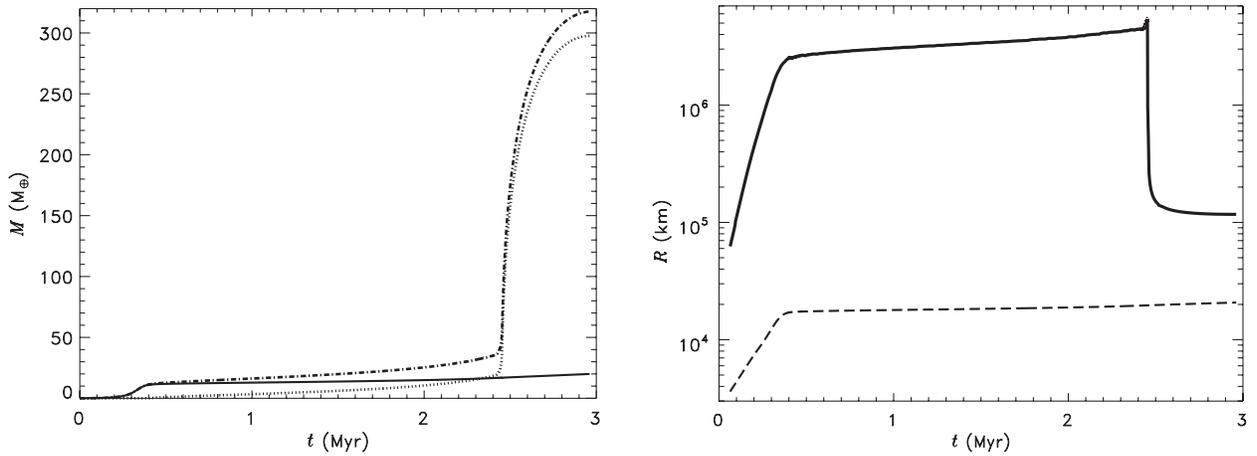

**Figure 1.** Formation models of a giant planet by core nucleated accretion need to take into account many physical effects. These include: *i*) Calculation of the planetesimal accretion rate onto the protoplanet; *ii*) Calculation of the interaction via gas drag of impacting planetesimals with the protoplanet's gaseous envelope; *iii*) Thermodynamics calculation of the envelope structure; *iv*) Calculation of the gas accretion rate from the protoplanetary disk onto the protoplanet during the phase of slow envelope growth; *v*) Hydrodynamics calculation of constraints on envelope size and disk-limited gas accretion rates during the phase of runaway gas accretion. Some results from one such calculation are reported here for a model of Jupiter's formation (*Lissauer et al.*, 2009). This model assumes that the planetesimal disk has $\Sigma_s \approx 10\,\mathrm{g\,cm^{-2}}$ and the grain opacity in the protoplanet's envelope is 2% the value of the interstellar medium. The gaseous disk is assumed to dissipate within 3 Myr. The left panel shows the mass of solids (solid line), of gas (dotted line), and the total mass of the planet (dot-dashed line) as functions of time. The right panel displays the radius of the planet (solid line) and that of the planet's solid core (dashed line) from the same model.

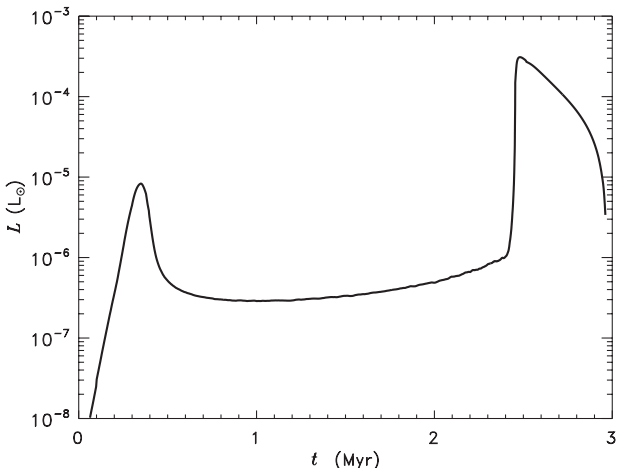

**Figure 2.** Luminosity as a function of time from the model reported in Figure 1. From the time of the first luminosity peak ($t \approx 0.35\,\mathrm{Myr}$) to the middle of the planetesimal isolation phase ($t \sim 1.6\,\mathrm{Myr}$), the luminosity of the protoplanet decreases by more than a factor of 25. This is mainly due to the reduced accretion rate of solids.

It thus appears that the duration of the slow contraction phase represents the main uncertainty in determining whether a giant planet can form before the gaseous component of the disk dissipates (i.e., within a few to several millions of years; see the chapter by Roberge and Kamp). However, it offers a natural explanation for the formation of planets such as Uranus and Neptune in our solar system, which do not posses very massive envelopes. These planets may have not been able, because of core and envelope conditions, or may have not had enough time, because of a dissipating disk, to acquire a large envelope. In addition, static atmospheric models indicate that the removal of grains by growth and settling from the radiative zone of a planet's envelope may significantly reduce their contribution to the opacity (*Movshovitz and Podolak*, 2008), hence shortening the contraction timescale of the envelope.

Quasi-static calculations of the envelope thermal structure suggest that the growth timescale of a protoplanet's envelope $\tau_e = M(dM_e/dt)^{-1}$ can be cast in the following parametric form (e.g., *Ida and Lin*, 2008)

$$\tau_e = \bar{\tau}_e \left(\frac{M_\oplus}{M}\right)^\xi, \qquad (15)$$

in which $M = M_c + M_e$ is the total planet's mass and $M_c \gtrsim M_{\mathrm{iso}}$. The timescale $\bar{\tau}_e$ depends on several factors, but mostly on the opacity of the envelope. For the model reported in Figure 1, $\bar{\tau}_e \sim 10^{10}$ years and $\xi \sim 3$. Notice, though, that many factors affect the values of parameters $\xi$ and, especially, of $\bar{\tau}_e$. In general, applicability of equation (15) requires some previous knowledge of the en-velope thermal conditions, for example through theoretical arguments or numerical modeling. When $M_e \gtrsim M_c$, the inner (densest) parts of the envelope, which contain most of the mass, effectively contribute to the compression of the outer parts. Yet, around this stage, the envelope may become gravitationally unstable and collapse (see Figure 1).

## 2.5. Critical Core Mass and Envelope Collapse

In order to gain some insight into the conditions leading to envelope collapse, we can construct simplified analytical models that qualitatively describe the physical state of the envelope. The first assumption we shall make is that a slowly contracting envelope evolves through stages of quasihydrostatic equilibria. Early studies of a static envelope surrounding a planetary core (e.g., *Perri and Cameron*, 1974) suggested that there may not be any static structure beyond some value of the core mass. Past such a *critical core mass*, the extended outer portions of the gaseous envelope become hydrodynamically unstable and collapse onto the core. Simplified models of hydrostatic envelopes can shed some light over the concept of critical core mass. One such model was used by *Stevenson* (1982) to interpret results of static envelope calculations performed by *Mizuno* (1980).

Let us assume that a gaseous envelope of mass $M_e$ surrounds a core of mass $M_c$ and that the envelope is spherically symmetric around the core. Hydrostatic equilibrium in an envelope shell of radius $r$ and thickness $dr$ implies that there is a balance between gravitational force and pressure acting at the shell surface boundaries. Therefore, the condition for hydrostatic equilibrium reads

$$\frac{dp}{dr} = -\frac{G\,m\,\rho}{r^2}, \qquad (16)$$

where $p$ and $\rho$ are, respectively, the pressure and density at radius $r$, and $m$ is the mass within $r$.

An important assumption we shall make is that energy is transported through the envelope via radiative diffusion (i.e., this model applies to a radiative envelope) and that the envelope is stable against convection. In general, this may be true only in the outer layers of the envelope (*Bodenheimer and Pollack*, 1986).

The amount of radiation energy transported through an envelope shell per unit surface area and unit time, i.e., the radiation flux, is $F_{\rm rad} = -\mathcal{D}dE_{\rm rad}/dr$, where the radiation energy density $E_{\rm rad} \propto T^4$ and $T$ is the temperature. The radiation diffusion coefficient, $\mathcal{D}$, is proportional to the speed of light times the mean free path of photons, $1/(\bar\kappa \rho)$, where $\bar\kappa$ is a frequency-integrated opacity (or mass absorption coefficient). Hence, for an envelope shell of mass $dm$, the radiation flux is also equal to (e.g., *Kippenhahn and Weigert*, 1990) $F_{\rm rad} = -16\pi r^2 \sigma_{\rm rad}/(3\bar\kappa)(dT^4/dm)$, where $\sigma_{\rm rad}$ is the Stefan-Boltzmann constant. If the luminosity through the envelope, $L$, is nearly constant then one can write $F_{\rm rad} = 4\pi r^2 L$, and thus

$$\frac{dT^4}{dm} = -\frac{3\,\bar\kappa\,L}{64\pi^2\,\sigma_{\rm rad}\,r^4}. \qquad (17)$$

Another assumption we shall rely upon is that the gradient $\nabla = d\ln T/d\ln p$ is nearly constant in the envelope. The temperature gradient can thus be cast in the form

$$\frac{dT}{dr} = \left(\frac{T}{p}\right)\left(\frac{dp}{dr}\right)\nabla. \qquad (18)$$

In order for energy to be transported via radiation alone, the gradient $\nabla$ must be smaller than its adiabatic value, $\nabla_{\rm ad}$. For a perfect monoatomic gas and negligible radiation pressure, $\nabla_{\rm ad} = 2/5$. If $\nabla \gtrsim \nabla_{\rm ad}$, the envelope becomes convectively unstable. By substituting equation (16) in place of the pressure gradient in equation (18) and using the equation of state for a perfect gas $p/\rho = \mathcal{R}T/\mu$ ($\mathcal{R}$ is the gas constant), we have that $dT/dr \sim -\mu G M \nabla/(\mathcal{R}r^2)$, where $M = M_c + M_e$, and hence $T \sim \mu G M \nabla/(\mathcal{R}r)$. In the above relations, $\mu$ indicates the mean-molecular weight, that is the average number of atomic mass units per gas particle. Notice that the approximation $m \sim M$ made above determines how deep in the envelope this temperature stratification is applicable.

Equation (17) can be written as

$$\frac{dm}{dr} = -\frac{4^4 \pi^2\,\sigma_{\rm rad}}{3\,\bar\kappa\,L}\,r^4\,T^3\frac{dT}{dr}, \qquad (19)$$

from which it follows that

$$\frac{dm}{dr} = \frac{4^4 \pi^2\,\sigma_{\rm rad}}{3\,\bar\kappa\,L}\left(\frac{\mu G M \nabla}{\mathcal{R}}\right)^4\frac{1}{r}. \qquad (20)$$

Integrating both sides of this equation along $r$, from the core radius, $R_c$, to the envelope radius, $R$, and neglecting opacity variations[4], one obtains

$$M - M_c = \mathcal{A}_0 \frac{(\mu\nabla)^4}{\bar\kappa\,L}\,M^4, \qquad (21)$$

in which $\mathcal{A}_0 = (\pi^2 \sigma_{\rm rad}/3)(4G/\mathcal{R})^4 \ln(R/R_c)$.

We seek for the largest value of the core mass that can have a fully radiative envelope in hydrostatic equilibrium. This can be obtained from equation (21), which gives an explicit function $M_c = M_c(M)$, if $L$ is independent of both $M$ and $M_c$. The value of $M$ for which $M_c$ is maximum is given by

$$M^{\rm cr} = \left[\frac{1}{4\,\mathcal{A}_0}\frac{\bar\kappa\,L}{(\mu\nabla)^4}\right]^{1/3}. \qquad (22)$$

Physically significant solutions of equation (21) can only be obtained for $M/M^{\rm cr} < 4^{1/3}$, and therefore the mass of *strictly* hydrostatic envelopes does not exceed $4^{1/3}\,M^{\rm cr}$. The critical core mass for radiative envelopes is found by substituting equation (22) in equation (21), which gives

$$M_c^{\rm cr} = \frac{3}{4}M^{\rm cr}. \qquad (23)$$

---

[4]A more general opacity law, given by the product of some power of $p$ and some power of $T$, would result in the same relation as equation (21), but with a modified form of coefficient $\mathcal{A}_0$ (*Wuchterl et al.*, 2000).

According to these simple analytical arguments, the critical core mass, $M_c^{\rm cr}$, is independent of the local density and temperature of the protoplanetary disk, it is weakly dependent on envelope opacity and luminosity, but more strongly dependent on $\mu$ and $\nabla$. Using radii and luminosity values before collapse from Figures 1 and 2, $\mu = 2.25$, $\nabla = 0.235$ (*Kippenhahn and Weigert*, 1990), and $\bar{\kappa} = 0.02\,{\rm cm}^2\,{\rm g}^{-1}$, one obtains $M_c^{\rm cr} \sim 16\,M_\oplus$ and a total mass of $\sim 21\,M_\oplus$ (recall that these values would apply to a strictly hydrostatic and fully radiative envelope).

As mentioned in Section 2.4, most of the protoplanet's luminosity during the slow growth phase may be due to gravitational energy released by accreted planetesimals. In these circumstances the luminosity can be approximated as $L \sim (GM_c/R_c)(dM_c/dt)$, in which $dM_c/dt$ is the planetesimal accretion rate, thus $L \propto M_c^{2/3}(dM_c/dt)$. Substituting in equation (21), one obtains $M_c$ as an implicit function of $M$. By differentiation of this function, one can readily show that again $M_c^{\rm cr} = (3/4)M^{\rm cr}$.

The deepest and densest layers of an envelope are typically convective (e.g., *Bodenheimer and Pollack*, 1986). Simple analytical models of hydrostatic envelopes can also be constructed under the hypothesis that energy is transported only via convection, as that proposed by *Wuchterl* (1993). By employing the adiabatic gradient $\nabla_{\rm ad}$, the temperature gradient becomes $dT/dr \sim -\mu GM\nabla_{\rm ad}/(\mathcal{R}r^2)$. Instead of an energy equation, in a fully convective envelope one can use the polytropic law $T/T_D = (\rho/\rho_D)^\Gamma$, in which $T_D$ and $\rho_D$ indicate values where the envelope merges with the disk. In a perfect monoatomic gas with negligible radiation pressure, $\Gamma = (5/3)\nabla_{\rm ad}$. By differentiation of the polytropic law, it follows that

$$\frac{dT}{d\rho} = \Gamma \left(\frac{T_D}{\rho_D}\right)\left(\frac{\rho}{\rho_D}\right)^{(\Gamma-1)}, \qquad (24)$$

which combined with the temperature gradient above yields

$$\Gamma\left(\frac{T_D}{\rho_D}\right)\left(\frac{\rho}{\rho_D}\right)^{(\Gamma-1)}\frac{d\rho}{dr} = -\left(\frac{\mu GM\nabla_{\rm ad}}{\mathcal{R}r^2}\right). \qquad (25)$$

Equation (25) can be integrated in radius to obtain the envelope's density stratification

$$\left(\frac{\rho}{\rho_D}\right)^\Gamma = -\left(\frac{\mu\nabla_{\rm ad}}{\mathcal{R}T_D}\right)\left(\frac{GM}{r}\right). \qquad (26)$$

The approximation $m \sim M$, introduced above to derive the temperature gradient, again determines the envelope depths to which this density stratification applies. In equation (26) the constant of integration is set to zero by choosing as outer envelope radius $R = \mu GM\nabla_{\rm ad}/(\mathcal{R}T_D)$. By integrating the quantity $4\pi r^2 \rho\, dr$ from the core radius, $R_c$, to $R$ we get

$$M - M_c = \mathcal{B}_0\, \rho_D \left(\frac{\Gamma}{3\Gamma - 1}\right)\left(\frac{\mu\nabla_{\rm ad}}{T_D}\right)^3 M^3, \qquad (27)$$

where $\mathcal{B}_0 = 4\pi(G/\mathcal{R})^3\left[1 - (R_c/R)^{(3-1/\Gamma)}\right]$, which can be approximated to $4\pi(G/\mathcal{R})^3$ since $R_c \ll R$ and $3 - 1/\Gamma > 1$ when radiation pressure can be ignored.

The largest core mass that bears a fully convective envelope in hydrostatic equilibrium corresponds to a total mass

$$M^{\rm cr} = \left[\frac{1}{12\pi}\left(\frac{3\Gamma - 1}{\Gamma}\right)\left(\frac{\mathcal{R}}{\mu G\nabla_{\rm ad}}\right)^3 \frac{T_D^3}{\rho_D}\right]^{1/2}. \qquad (28)$$

In order for equation (27) to admit physically significant solutions, $M/M^{\rm cr} < \sqrt{3}$. According to these simple arguments, strictly hydrostatic and convective envelopes are limited in mass to $\sqrt{3}\,M^{\rm cr}$. From equations (27) and (28), one finds that the critical core mass of a convective envelope is

$$M_c^{\rm cr} = \frac{2}{3}M^{\rm cr}. \qquad (29)$$

Unlike the solution for radiative envelopes, the critical core mass for convective envelopes explicitly depends on the disk density and temperature. Using values $\rho_D \sim 10^{-10}\,{\rm g\,cm}^{-3}$ and $T_D \sim 100\,{\rm K}$, which would apply to a protoplanetary disk at $\sim 5\,{\rm AU}$ from a solar-mass star if there was no planet, equation (28) gives $M^{\rm cr} \sim 170\,M_\oplus$. However, once $M$ achieves several tens of Earth masses, both $\rho_D$ and $T_D$ are affected by disk-planet interactions and hence depend on $M$. In these cases, because of the gravitational perturbation produced by the protoplanet, local disk densities as well as temperatures can be smaller than the corresponding values in absence of the protoplanet (*D'Angelo et al.*, 2003a), if effects of stellar irradiation can be neglected.

The two analytic solutions for fully radiative and fully convective envelopes can be combined to obtain an estimate of the critical core mass for composite convective-radiative envelopes. Following the same line of argument, we consider a two-layer model in which the "convective" solution applies to the inner convective layer that extends from the core radius, $R_c$, to the radius $R_{\rm clb}$, which plays the role of envelope radius ($R$) in the convective solution given above. Accordingly, temperature $T_{\rm clb}$ and density $\rho_{\rm clb}$ at the convective layer boundary, $R_{\rm clb}$, replace the disk values $T_D$ and $\rho_D$, respectively. Then, indicating with $M_{\rm clb}$ the mass within $R_{\rm clb}$ (core mass plus mass of the convective layer), equation (29) gives a critical value $M_{\rm clb}^{\rm cr} = (3/2)M_c^{\rm cr}$.

The "radiative" solution applies to the outer radiative layer, whose boundaries are at the convective layer (outer) radius $R_{\rm clb}$, which plays the role of core radius ($R_c$) in the radiative solution above, and at the envelope radius $R$. Thus, the mass within the convective layer boundary, $M_{\rm clb}$, replaces the core mass ($M_c$) in the radiative solution above. Equation (23) gives the critical value of the total mass $M^{\rm cr} = (4/3)M_{\rm clb}^{\rm cr}$. Therefore, the largest planet mass for which both the convective and radiative layers, and hence the entire envelope, can be in hydrostatic equilibrium is

$$M^{\rm cr} = \frac{4}{3}M_{\rm clb}^{\rm cr} = 2M_c^{\rm cr}. \qquad (30)$$

According to equation (30), a composite convective-radiative envelope collapses once the envelope mass is equal to the core mass. This prediction of the simple analytic model is in agreement with the results from the detailed calculation shown in the left panel of Figure 1. Equation (22) can still be used to estimate $M^{\rm cr}$, where now $\mathcal{A}_0 \propto \ln(R/R_{\rm clb})$. At the epoch of envelope collapse, the calculation in Figure 1 gives a ratio $R/R_{\rm clb} \approx 4.74$ which, together with the other values adopted above, yields $M^{\rm cr} \sim 32\,M_\oplus$ and $M_c^{\rm cr} \sim 16\,M_\oplus$, consistent with the results illustrated in the Figure's left panel.

## 2.6. Disk-Limited Gas Accretion Rates

During the collapse phase, large amounts of gas can be accreted from the protoplanetary disk. The rate at which the protoplanet's envelope grows soon becomes very large. When the total mass is $M \sim 20\,M_\oplus$, equation (15) yields an envelope growth timescale of order $10^6$ years. But once the envelope mass, $M_e$, exceeds the core mass (equation (30)), this length of time can shorten by orders of magnitude. In the model illustrated in Figure 1, when $M_e \sim 2\,M_c$ the growth timescale becomes of order $10^4$ years.

The protoplanetary disk feeds gas to the planet's vicinity. Yet, the rate at which such gas supply occurs is limited. In this stage, since a protoplanet can basically accrete gas at arbitrarily large rates, gas accretion is governed by hydrodynamical factors that involve tidal (i.e., gravitational) interactions between the growing planet and the disk . To derive simple estimates of disk-limited gas accretion rates[5], $dM/dt$, we can assume that a gas parcel orbits about the star, at a distance $a + \Delta a$, with an orbital velocity relative to the planet of order $\Omega \Delta a$ (see Section 2.2), where $\Omega$ is the angular velocity of the planet and $a$ its orbital radius. The protoplanet is able to capture gas within an effective radius $R_{\rm gc}$ (i.e., $\pi R_{\rm gc}^2$ is its effective cross-section for gas capture) smaller than the disk's local thickness, $H$. If $\rho$ is the gas volume density, then the mass flux through the planet's cross-section is $\rho \Omega R_{\rm gc}$. The gas accretion rate is given by the mass flux times the planet's cross-section, that is

$$\frac{dM}{dt} \sim \frac{\Sigma}{H}\,\Omega\,R_{\rm gc}^3\,,\qquad(31)$$

in which the volume density is expressed in terms of surface density ($\rho \sim \Sigma/H$).

There are two relevant length-scales over which a protoplanet can attract gas: the Hill and Bondi radii. The Hill radius, $R_{\rm H}$, is defined in equation (10) (in which $M_s$ is replaced with the total mass $M$) and represents a measure of the distance past which centrifugal forces and gravitational forces by the star dominate over the gravitational force exerted on the gas by the planet. The Bondi radius is

$$R_{\rm B} = \frac{G\,M}{c^2}\qquad(32)$$

---

[5] In this phase, we designate the accretion rate of gas simply as $dM/dt$ since total mass variations are overwhelmingly determined by accretion of gas, regardless of whether there is continued accretion of solids.

($c = H\Omega$ is the gas sound speed) and marks the distance beyond which the thermal energy of the gas is larger than the gravitational energy binding that gas to the planet. In the case that $R_{\rm B} < R_{\rm H}$, forces due to pressure gradients within the Hill sphere prevent gas from becoming bound to the protoplanet. Then, the Bondi radius is the relevant distance over which gas can be accreted. We will refer to this regime as *Bondi-type gas accretion*. The Hill radius becomes the relevant distance for gas accretion if $R_{\rm H} < R_{\rm B}$, which will be designated as *Hill-type gas accretion* regime. Therefore, in the general case, the effective radius for gas capture of a protoplanet embedded in a gaseous disk is

$$R_{\rm gc} = \min(R_{\rm B}, R_{\rm H})\,.\qquad(33)$$

By substituting equation (32) in equation (31), one finds that the protoplanet's growth rate, $(dM/dt)/M$, in the Bondi-type gas accretion regime is

$$\frac{1}{\tau_{\rm B}} = \mathcal{C}_{\rm B} \Omega \left(\frac{a^2\Sigma}{M_\star}\right)\left(\frac{a}{H}\right)^7\left(\frac{M}{M_\star}\right)^2,\qquad(34)$$

whereas, in the Hill-type gas accretion regime, the growth rate is found by using equation (10) (with $M$ in place of $M_s$) and equation (31), resulting in

$$\frac{1}{\tau_{\rm H}} = \frac{1}{3}\mathcal{C}_{\rm H} \Omega \left(\frac{a^2\Sigma}{M_\star}\right)\left(\frac{a}{H}\right).\qquad(35)$$

The dimensionless constants $\mathcal{C}_{\rm B}$ and $\mathcal{C}_{\rm H}$ account for small corrections due to effects neglected in deriving this simple model and should be of order unity. They can be obtained from direct numerical simulations. Notice that $dM/dt \propto M^3$ for Bondi-type gas accretion (implying a faster than exponential growth) while $dM/dt \propto M$ for Hill-type gas accretion (implying an exponential growth).

Bondi-type gas accretion is a very steep function of the disk thickness, i.e., of the local gas sound speed. If the disk is locally very warm, high thermal energy prevents gas from becoming bound and hence accreted (the planet's cross-section for gas accretion decreases for increasing temperature). Hill-type gas accretion is less sensitive on disk temperature because in this regime gravitational factors dominate over thermal ones.

In general, the mass growth rate of the protoplanet, $1/\tau_D$, which is limited by disk hydrodynamics, is given by

$$\frac{1}{\tau_D} = \begin{cases} 1/\tau_{\rm B} & \text{for } M < M_{\rm tr} \\ 1/\tau_{\rm H} & \text{for } M \geq M_{\rm tr} \end{cases},\qquad(36)$$

where the transition between the two regimes occurs at the transition mass $M_{\rm tr}$, for which $\tau_{\rm H} = \tau_{\rm B}$, and thus

$$M_{\rm tr} = \frac{M_\star}{\sqrt{3}}\sqrt{\frac{\mathcal{C}_{\rm H}}{\mathcal{C}_{\rm B}}}\left(\frac{H}{a}\right)^3.\qquad(37)$$

This simple gas accretion model can be compared against results from three-dimensional hydrodynamical simulations of a planet gravitationally interacting with a disk and growing in mass at a disk-limited gas accretion rate. Figure 3 (left panel) shows numerical results (solid line) and the growth rate given by equation (36) (dashed line). Agreement is found for values of the constants $\mathcal{C}_B \simeq 2.6$ and $\mathcal{C}_H \simeq 0.9$. Note that the mass scaling, i.e., the slopes of the two dashed line segments, is correctly predicted by the simple model and so is the transition mass between the two regimes ($M_{\mathrm{tr}} \approx 4\times 10^{-5}\,M_\star$, or $\approx 14\,M_\oplus$ if $M_\star = 1\,M_\odot$). The disk thickness in the example shown in Figure 3 is $H/a = 0.05$, which corresponds, for typical disk properties, to a local temperature $T \sim 100\,\mathrm{K}$. Since the gas sound speed $c \propto \sqrt{T}$, a factor of 2 increase in the local temperature would reduce the planet's growth rate by an order of magnitude during the Bondi-type accretion regime.

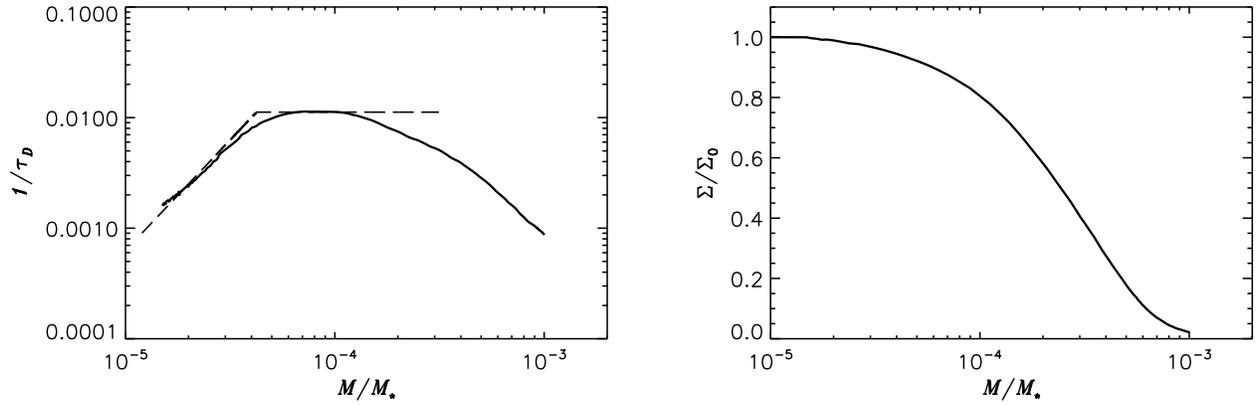

**Figure 3.** *Left:* Mass growth rate, $1/\tau_D = (dM/dt)/M$, versus mass of a protoplanet accreting gas at a limiting rate provided by a protoplanetary disk. These results were obtained from a three-dimensional, high-resolution hydrodynamical model of a growing protoplanet interacting with a disk of initial surface density at $a = 5.2\,\mathrm{AU}$ of $100\,\mathrm{g\,cm^{-2}}$, relative thickness $H/a = 0.05$, and turbulence parameter $\alpha = 4\times 10^{-3}$ (*D'Angelo and Lubow*, 2008). The timescale $\tau_D$ is in units of orbital periods. The dashed line represents the growth rate given by equation (36): the slanted portion corresponds to the rate in the Bondi-type gas accretion regime ($1/\tau_B$ with $\mathcal{C}_B \simeq 2.6$) and the horizontal portion corresponds to the Hill-type gas accretion rate ($1/\tau_H$ with $\mathcal{C}_H \simeq 0.9$). At large planet masses, the growth rates drop due to the formation of the tidally produced gap in the local density distribution. *Right:* Average surface density distribution of the disk's gas, $\Sigma$, near the planet relative to the local unperturbed (i.e., without a planet) value, $\Sigma_0$, as a function of the planet's mass.

As the planet's mass increases, the effective radius for gas capture may exceed the disk thickness. Moreover, density perturbations due to tidal interactions of the protoplanet with the disk are no longer negligible. The gas density along the planet's orbit starts to be affected (see right panel of Figure 3), and the simple accretion model becomes inapplicable beyond some value of the protoplanet mass. In the example shown in Figure 3, this occurs for masses $M \gtrsim 10^{-4}\,M_\star$, when disk-planet interactions have changed the local surface density by more than 20% of its unperturbed value (i.e., when there is no planet in the disk), as illustrated in the right panel.

### 2.7. Gap Formation

A protoplanet embedded in a disk exerts a gravitational torque on the gas interior of $a - |\Delta a|$ and exterior of $a + |\Delta a|$ whose magnitude is of order

$$\mathcal{T}_g \approx \frac{\Sigma}{f} a^4 \Omega^2 \left(\frac{a}{\Delta a}\right)^3 \left(\frac{M}{M_\star}\right)^2, \quad (38)$$

which leads to exchange of orbital angular momentum between the planet and the disk (see the chapter by *Lubow and Ida* for a derivation of this torque). The factor $f$ is typically of order unity. It can be shown that $|\Delta a|$ is the larger between the Hill radius, $R_H$, and the disk scale-height, $H$. The sign of $\mathcal{T}_g$ is positive for material lying outside of the planet's orbit ($\Delta a > 0$) and negative for material inside of it ($\Delta a < 0$). Hence, material orbiting outside of the planet's orbit gains angular momentum, moving towards larger radii, whereas material inside of the planet's orbit loses angular momentum, moving towards smaller radii. This process tends to deplete the disk of gas along the planet's orbit, thus forming an annular gap in the local density distribution (e.g., *Lin and Papaloizou*, 1986). However, in a viscous (Keplerian) disk, because of differential rotation gas is also subject to a viscous torque generated by viscous friction between adjacent disk rings. The viscous torque exerted by material inside of the orbital radius $a$ on material outside of $a$ is (*Lynden-Bell and Pringle*, 1974)

$$\mathcal{T}_v \approx 3\pi \nu\, a^2\, \Sigma\, \Omega. \quad (39)$$

This torque tends to smooth out density gradients and redistribute material across the planet's orbit, filling in the gap.

A condition for gap formation requires a net gain of orbital angular momentum for material outside the radius $a$, that is $\mathcal{T}_g - \mathcal{T}_v > 0$, and a net loss of orbital angular momentum for material inside $a$, $\mathcal{T}_g + \mathcal{T}_v < 0$. If $H \geq R_H$, the condition $|\mathcal{T}_g| > \mathcal{T}_v$ translates into a condition for the minimum planet's mass necessary to open up a density gap in a disk

$$\left(\frac{M}{M_\star}\right)^2 \gtrsim 3\pi f \alpha \left(\frac{H}{a}\right)^5, \quad (40)$$

where the turbulence parameter $\alpha$ is defined in Section 2.1. In case of very cold disks or very massive planets $R_H > H$ and the torque inequality gives $(M/M_\star) \gtrsim \pi f \alpha (H/a)^2$. For conditions adopted in the simulations reported in Figure 3, substantial gas depletion is expected for $M \gtrsim 2 \times 10^{-4} M_\star$ (or $M \gtrsim 60 M_\oplus$ for a disk around a solar-mass star). In accord with this estimate, the right panel of the Figure indicates that the average surface density along the planet's orbit is reduced by 40% of its unperturbed value when $M \gtrsim 2 \times 10^{-4} M_\star$.

The timescale for gap formation can be estimated from the equation of motion of a thin viscous Keplerian disk, which also evolves under the action of gravitational torques exerted by an embedded object (*Lin and Papaloizou*, 1986)

$$\pi a \frac{\partial \Sigma}{\partial t} = \frac{\partial}{\partial a}\left[\frac{1}{a\Omega}\frac{\partial}{\partial a}(\mathcal{T}_v - \mathcal{T}_g)\right], \quad (41)$$

where radius $a$ is now interpreted as a variable indicating the distance from the star. Equation (41) can be derived from imposing conservation of mass and angular momentum in an axisymmetric and flat disk. Ignoring viscous torques ($\mathcal{T}_v < |\mathcal{T}_g|$) and taking the approximation $\partial/\partial a \to 1/\Delta a$ in equation (41), defining the gap formation timescale as $\tau_{\text{gap}} = \Sigma|\partial\Sigma/\partial t|^{-1}$ and using equation (38), it follows that

$$\Omega \tau_{\text{gap}} \sim \pi f \left(\frac{M_\star}{M}\right)^2 \left(\frac{|\Delta a|}{a}\right)^5, \quad (42)$$

with $|\Delta a|$ equal to the larger of $H$ and $R_H$. This timescale is typically short, on the order of tens of orbital periods.

Once gap formation starts, accretion rates drop because of gas depletion around the planet. According to condition (40), lower disk viscosity allows for gas depletion around smaller mass planets. By using equation (37), one finds that for

$$\alpha \lesssim \frac{\mathcal{C}_H}{9\pi f \mathcal{C}_B}\left(\frac{H}{a}\right), \quad (43)$$

gas depletion begins before the transition to a Hill-type gas accretion and the growth rate starts to decline for smaller mass planets. In a disk of thickness $H/a \approx 0.05$, this may occur for $\alpha$ less than few times $10^{-4}$.

Effects of gap formation can be seen in Figure 4. The Figure shows disk-limited gas accretion rates, derived from hydrodynamical simulations (see *D'Angelo et al.*, 2003b, for details), for protoplanets embedded in a moderately viscous ($\alpha = 4 \times 10^{-3}$) and in a low-viscosity ($\alpha = 4 \times 10^{-4}$) disk. Notice, though, that if $\tau_{\text{gap}} > \tau_H$, i.e., $f \mathcal{C}_H (a^2 \Sigma/M_\star) \gtrsim (M/M_\star)^2 (a/H)^4$, Hill-type gas accretion may persist until $R_H \sim H$.

Disk-limited gas accretion rates can be quite large, unless the disk's gas density is low. A Saturn-mass planet orbiting a solar-mass star at $\sim 5$ AU in a disk, whose gas surface density just outside the gap region is of order $100 \text{ g cm}^{-2}$, may accrete gas at rates $\sim 10^{-3}$–$10^{-2} M_\oplus$ per year, thereby reaching a Jupiter's mass in $10^4$–$10^5$ years.

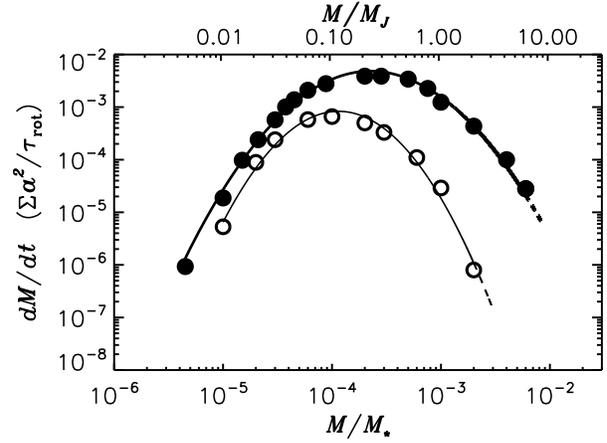

**Figure 4.** Disk-limited gas accretion rates as a function of the planet mass obtained from three-dimensional, high-resolution hydrodynamical calculations of a planet interacting with a protoplanetary disk. Accretion rates are in units of the unperturbed surface density at the planet's orbital radius, $a$, and the planet's orbital period, $\tau_{\text{rot}}$. The top axis uses $M_\star = 1 M_\odot$. Filled circles correspond to results for a disk with a turbulence parameter $\alpha = 4 \times 10^{-3}$ (see Section 2.1). Empty circles are for a disk with $\alpha = 4 \times 10^{-4}$. The disk's aspect ratio is $H/a = 0.05$. The curves represent fits to the data.

### 2.8. Final Masses of Giant Planets

It is not yet entirely clear what all the factors are that determine the final mass of a giant planet. Yet, tidal truncation of the disk by gravitational interactions with the planet is likely one of the main factors. As indicated in the right panel of Figure 3, gaps can become quite deep around a Jupiter-mass planet, even at moderate values of the turbulence parameter $\alpha$. However, gas can filter through the tidal barrier (i.e., the gap) and continue to accrete towards the planet.

This can be seen in Figure 5, which shows the gap in the surface density distribution around a giant planet together with trajectories of material moving towards the planet along the inner and outer edges of the gap. The trajectories are drawn in a frame co-moving with the planet. This material becomes trapped in the inner parts of the planet's Roche lobe and is eventually accreted. As mentioned in the previous section, accretion through a gap can be quite efficient. Assuming a gas surface density of $\sim 100 \text{ g cm}^{-2}$ just outside the gap region at $\sim 5$ AU, the growth timescale (due to disk-limited gas accretion), $\tau_D = M(dM/dt)^{-1}$, of a Jupiter-mass planet in a disk with $\alpha \sim 10^{-3}$ and $H/a \approx 0.05$ would be several thousands of orbital periods (see Figure 4). Around a solar-mass star, the tidal barrier alone would constrain the planet's final mass to about 6–7 $M_J$.

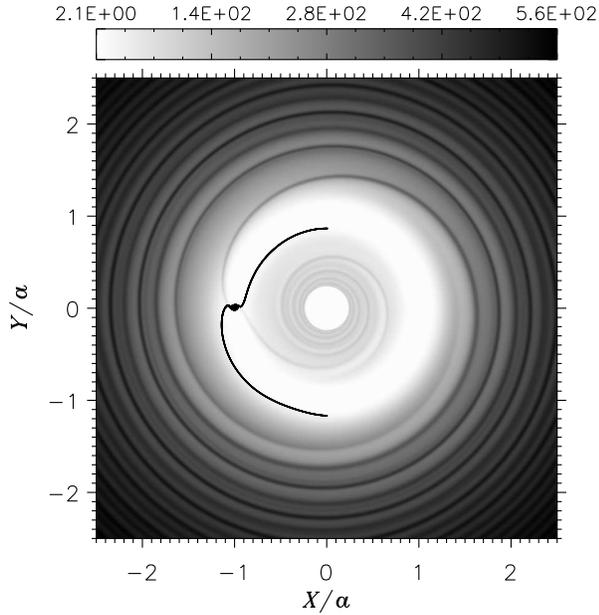

**Figure 5.** Surface density of a gaseous disk containing a protoplanet whose mass is $M = 10^{-3} M_\star$ (i.e., one Jupiter's mass for a solar-mass star). The relative disk thickness is $H/a = 0.05$ and the turbulence parameter is $\alpha = 4 \times 10^{-3}$. The planet is located at $(-1, 0)$, the star at $(0, 0)$, and the disk is rotating in the counterclockwise direction. The greyscale bar is in cgs units. The plot shows the density gap along the planet's orbit and the wave pattern generated by disk-planet interactions. The two black lines represent trajectories of gas in the co-moving frame of the planet. These gas parcels move along the inner and outer gap edges, become gravitationally bound to the planet, and are eventually accreted.

A lower disk temperature, i.e., a smaller disk thickness, can increase the planet's tidal barrier and reduce gas accretion. However, it is unclear whether accretion can be stopped around $M \sim 1\, M_J$ for reasonable values of $H/a$ (*Lubow and D'Angelo*, 2006). Lower disk viscosity also helps reduce disk-limited gas accretion rates, as indicated in Figure 4. Assuming a disk around a solar-mass star with $\alpha \sim 10^{-4}$, for the same example made above, the tidal barrier would limit the planet's final mass to $\sim 2\, M_J$.

The considerations above do not take into account the fact the a protoplanetary disk evaporates due to irradiation from the central star and, possibly, from other external sources (*Hollenbach et al.*, 2000). Both observations and theory suggest that photoevaporation timescales are on the order of a few million years. Therefore, disk consumption by photoevaporation may become a relevant process in the final stages of a giant planet's growth. It is likely that a combination of these three factors, i.e., a reducing disk thickness $H/a$, a decaying turbulence $\alpha$, and gas depletion by photoevaporation, all of which may take place as a protoplanetary disk ages, plays a significant role in determining final masses of gas giant planets.

Once gas accretion stops, a giant planet evolves at a constant mass. The planet gradually cools down and contracts on a Kelvin-Helmholtz timescale (equation (14)). The luminosity is mostly due to gravitational energy released in the contraction.

### 2.9. Orbital Migration of Protoplanets

Gravitational interactions between a protoplanetary disk and an embedded planet result in an exchange of orbital angular momentum (see the chapter by *Lubow and Ida* for a thorough discussion on this topic). Assuming that $H \gtrsim R_\mathrm{H}$ and that condition (40) for gap formation is not satisfied, the disk's surface density is largely unperturbed and equation (38) applies.

If the planet exerts a torque $\mathcal{T}_g$ on the gas outside its orbit, then conservation of angular momentum dictates that this gas exerts the same torque (with an opposite sign) on the planet. Therefore, the planet loses orbital angular momentum to material exterior to its orbit. By the same principle, the planet gains orbital angular momentum from material interior to its orbit. The torque expression in equation (38) is symmetrical with respect to the planet's radial position, which would result in a zero net torque acting on the planet. However, one can show that, because of global variations of the disk properties across the planet's orbital radius $a$, the net torque exerted on the planet by disk material is of order $|\mathcal{T}_p| \sim (H/a) |\mathcal{T}_g|$ (*Goldreich and Tremaine*, 1980; *Ward*, 1997), hence

$$\mathcal{T}_p \approx \mathcal{C}_\mathrm{I} \Sigma a^4 \Omega^2 \left(\frac{a}{H}\right)^2 \left(\frac{M}{M_\star}\right)^2. \quad (44)$$

Quantity $\mathcal{C}_\mathrm{I}$ depends on the radial gradients of surface density, temperature, and pressure of the disk across the planet's orbit. In typical circumstances, this quantity is negative and of order unity (*Tanaka et al.*, 2002).

The orbital angular momentum of a planet on a circular orbit is $L_p = M a^2 \Omega$. If the planet is acted upon by a torque $\mathcal{T}_p$, conservation of angular momentum imposes that $dL_p/dt - \mathcal{T}_p = 0$, and thus

$$\frac{1}{M}\frac{dM}{dt} + \frac{1}{2a}\frac{da}{dt} = \frac{\mathcal{T}_p}{M a^2 \Omega}. \quad (45)$$

If the first term on the lefthand side of equation (45) is negligible compared to the second, then $da/dt = 2\mathcal{T}_p/(Ma\Omega)$. This phenomenon is referred to as *orbital migration*. Since the torque experienced by the planet is generally negative, the orbit shrinks. When the torque acting on a (non-gap opening) planet is of the type in equation (44), migration is said of *type I*. The migration timescale, $a|da/dt|^{-1}$, in these circumstances is

$$\Omega \tau_\mathrm{I} \sim \left(\frac{M_\star}{a^2 \Sigma}\right)\left(\frac{M_\star}{M}\right)\left(\frac{H}{a}\right)^2. \quad (46)$$

Orbital decay due to type I migration can be quite rapid. A $10\, M_\oplus$ ($M = 3 \times 10^{-5} M_\star$) planetary core interacting with a disk, whose surface density is $\Sigma \sim 100\,\mathrm{g\,cm^{-2}}$ and $H/a \sim 0.05$, would migrate on a characteristic timescale of $\sim 4 \times 10^4$ orbital periods. For a planet orbiting at $\sim 5\,\mathrm{AU}$ around a solar-mass star this period would amount to several times $10^5$ years, a factor of a few shorter than the thermally-regulated envelope phase (see Section 2.4) shown in the left panel of Figure 1. Such rapid migration rates may be difficult to reconcile with formation via core accretion of giant planets that orbit at large distances from their stars, unless the time spent by a planet in the slow growth stage and the migration timescale $\tau_\mathrm{I}$ are actually closer (see, e.g., *Alibert et al.*, 2005).

Once a planet becomes a gas giant, a density gap forms along its orbit (because condition 40 is typically satisfied) where the surface density is drastically depleted (see right panel of Figures 3 and 5). In the limit of a very clean gap, there is a balance between viscous torques and gravitational torques at gap edges (interior and exterior of the planet's orbit) and the planet remains locked in the gap. This argument is valid as long as the disk inside the orbit of the planet is not significantly depleted. The planet drifts towards the star carried by the viscous diffusion of the disk, if the local mass of the disk is comparable to the planet's mass, or larger. This is referred to as *type II* migration. Viscous diffusion through the disk occurs on a timescale $a|da/dt|^{-1} \sim a^2/\nu$, and therefore

$$\Omega \tau_\mathrm{II} \sim \frac{1}{\alpha} \left(\frac{a}{H}\right)^2. \qquad (47)$$

In typical circumstances, there is residual gas in the gap region and there may not be an exact balance between viscous and gravitational torques at gap edges. Nonetheless, results from multi-dimensional simulations indicate that migration occurs on a timescale of order given by equation (47) in the presence of a sufficiently deep gap (e.g., *D'Angelo and Lubow*, 2008). In a disk with $\alpha \sim 10^{-3}$ and $H/a \sim 0.05$, a gap-opening planet would undergo orbital migration on timescales of many tens of thousands of orbital periods, or many $10^5$ years at $\sim 5\,\mathrm{AU}$ from a solar-mass star.

According to equation (47), $\tau_\mathrm{II} \propto a^{3/2}$ (if $H/a$ is approximately constant) and thus the migration timescale becomes shorter as a giant planet approaches the star. However, efficient exchange of orbital angular momentum between a gas giant and a protoplanetary disk requires that the local disk mass is at least comparable with the planet's mass, i.e., $\pi a^2 \Sigma \gtrsim M$. If this is not the case, planet's inertia acts to slow down migration. Strict type II migration of a $1\,M_J$ planet would therefore necessitate a gas surface density $\Sigma$ at both gap edges greater than $\sim 100\,\mathrm{g\,cm^{-2}}$ at $5\,\mathrm{AU}$ from the star and $\gtrsim 2.8 \times 10^3\,\mathrm{g\,cm^{-2}}$ at $1\,\mathrm{AU}$.

If the disk inside the planet's orbit is significantly depleted (due to the tidal barrier of a massive planet), conservation of angular momentum at the gap edge exterior of the planet's orbit requires that $dL_p/dt + \mathcal{T}_v = 0$ (e.g., *Syer and Clarke*, 1995), where the first term on the lefthand side is the rate of change of the planet's orbital angular momentum and the second term is the (positive) viscous torque exerted on the disk, given by equation (39). Thus, for a nearly-truncated disk at the gap edge exterior of the planet's orbit

$da/dt = -2\mathcal{T}_v/(Ma\Omega)$, and the migration timescale is

$$\Omega \tau_\mathrm{ED} \sim \frac{1}{6\pi\alpha} \left(\frac{a}{H}\right)^2 \left(\frac{M}{a^2 \Sigma}\right). \qquad (48)$$

Inertia effects are expected to be important at small orbital radii, when a disk is sufficiently depleted. Migration of giant planets provides a natural explanation for the occurrence of Jupiter-mass planets orbiting within a few tenth of AU from their parent stars, and whose existence would be otherwise difficult to explain by means of *in situ* formation via core accretion or disk instability (as we shall see in Section 3).

**2.10. Examples of Giant Planet Evolution Tracks**

An illustrative example of models combining planet's growth and orbital migration is displayed in Figure 6. These results are obtained from calculations that encapsulate in a simplified manner all the basic aspects of a core accretion model and standard migration theory discussed thus far. We consider here the case of a planetary embryo embedded in an axisymmetric (around the star) disk, which evolves under the action of viscous torques (equation (39)), gravitational torques, and photoevaporation induced by the central star. All three effects are accounted for in equation (41) once the lefthand side of that equation is replaced with $\pi a \partial (\Sigma + \Sigma_\mathrm{pe})/\partial t$, in which $\partial \Sigma_\mathrm{pe}/\partial t$ is the mass loss flux due to photoevaporation.

Tidal torques exerted by the planet on the disk are incorporated in equation (41) through a set of torque density distributions (torque per unit disk mass as a function of radius), which cover the relevant range of planet masses and are obtained from three-dimensional hydrodynamical simulations of disk-planet interactions (see *D'Angelo and Lubow*, 2008, for details). This procedure allows us to encompass the different regimes of orbital migration experienced by the planet as it grows and discussed in Section 2.9.

We use a simplified form of the mass loss rate from the disk that is zero within the "gravitational" radius $a_g \approx 10\,(M_\star/M_\odot)\,\mathrm{AU}$, where the thermal energy of the gas equals in magnitude its gravitational energy, otherwise $\partial\Sigma_\mathrm{pe}/\partial t = \dot{\Sigma}^g_\mathrm{pe}(a_g/a)^{5/2}$ (see the chapter by Roberge and Kamp). The mass loss flux at radius $a_g$ is $\dot{\Sigma}^g_\mathrm{pe} = 3.7 \times 10^{-13}\,(\Phi^{41}_\mathrm{pe})^{1/2}\,(M_\star/M_\odot)^{-3/2}$ in units of $M_\odot\,\mathrm{AU}^{-2}$ per year. The constant $\Phi^{41}_\mathrm{pe}$ is the rate of ionizing photons emitted by the star, $\Phi_\mathrm{pe}$, in units of $10^{41}\,\mathrm{s}^{-1}$.

At time $t = 0$, the disk contains $\approx 0.022\,M_\odot$ of gas within about $40\,\mathrm{AU}$ (*Davis*, 2005). The disk evolution equation is solved numerically, adopting a turbulence parameter $\alpha \approx 4 \times 10^{-3}$ and a disk scale-height $H/a = 0.05$. The evolution begins at $13\,\mathrm{AU}$ from a solar-mass star.

A planet embryo, of initial mass $0.1\,M_\oplus$, accretes solids at an oligarchic rate ($\propto M^{2/3}$, see Section 2.2) and becomes isolated when $M \approx 13\,M_\oplus$. The gas accretion rate, $dM/dt$, is given by the smaller of $M/\tau_e$ (equation (15)) and the upper fit to data in Figure 4. Three values of $\Phi^{41}_\mathrm{pe}$ are used: 1, 10, and 50. Thicker and thinner lines represent cases in which the planet's evolution starts at time $t = 0$ and 1.5 Myr, respectively. The evolution tracks illustrated in Figure 6 highlight the importance of the disk's initial conditions in which a giant planet's core begins to evolve and on the details of disk photoevaporation. It also supports the contention that early core accumulation may lead to substantial amounts of radial migration, whereas late-stage formation scenarios may imply much less orbital decay.

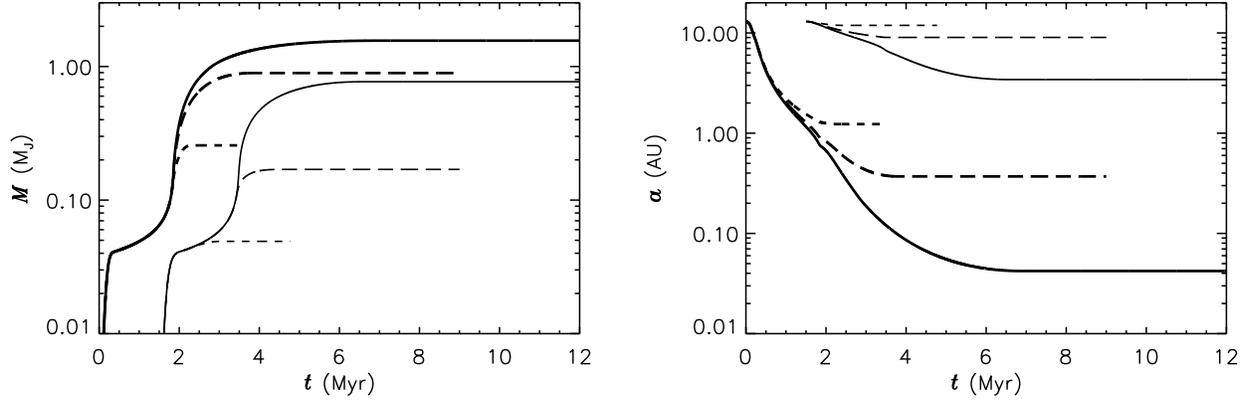

**Figure 6.** Evolution tracks of giant planet formation via core accretion obtained from models that account for orbital migration by tidal interactions with the protoplanetary disk (see Section 2.10 for details). The total mass (left) and orbital radius (right) of the planet are shown as a function of time. Two sets of tracks represent cases in which the planet evolution begins at time $t = 0$ (thicker lines) and at time $t = 1.5$ Myr (thinner lines). Within each set of tracks, three different line types represent different values of the applied rate of ionizing photons emitted by the star: $\Phi_{\mathrm{pe}} = 10^{41}\,\mathrm{s}^{-1}$ (solid lines), $10^{42}\,\mathrm{s}^{-1}$ (long-dashed lines), and $5 \times 10^{42}\,\mathrm{s}^{-1}$ (short-dashed lines).

### 2.11. Summary of Core Accretion Models

The formation of a gas giant via core nucleated accretion may take about one to a few Myr and is initiated by a solid planetary core of at least a few Earth masses. Core formation requires that the heavy elements of the disk have condensed out and coagulated into planetesimals. Although the process of planetesimal formation is still under scrutiny, small bodies of $\sim 10$ to $\sim 100$ km in size are observed in the Kuiper belt and Neptune-mass planets are observed around the Sun and other stars. Core accretion models provide an explanation for the existence not only of the four giant planets in the solar system, but also for the majority of giant planets that have been observed around other stars.

Giant planets contain large amounts of Hydrogen and Helium, which are acquired in gaseous form from protoplanetary disks. Therefore, gas dispersal timescales of disks set a strict upper limit on the time available to form these planets (a few to $\sim 10$ Myr). The phase of slow envelope contraction that precedes runaway gas accretion seems to represent the greatest hurdle to overcome since it may last longer than disk's gases, at least in some situations. Orbital decay by tidal torques may require formation in evolved rather than young disks to avoid amounts of radial migration that are too large. Additionally, since the oligarchic growth timescale is $\propto a^{3/2} M_\star^{-1/2}$, core formation at orbital distances of many tens of AU or around low-mass stars (e..g., red dwarfs) may take too long to produce a gas giant.

### 3. DISK INSTABILITY MODELS

Stars form through the gravitational collapse of interstellar clouds, and *Boss* (1997) has championed the idea (*Kuiper*, 1951; *Cameron*, 1978) that gas giant planets might form from protoplanetary disks in a similar way. This is commonly referred to as the *disk instability* theory of gas giant planet formation, because the mechanism by which the planets form is the onset of gas-phase *gravitational instabilities* (hereafter referred to as GIs) leading to *fragmentation* of a protoplanetary disk into bound, self-gravitating clumps. An important distinction with core accretion is that, if the conditions prevail for disk instability to work, then planets can form directly out of the gas phase within only a few to tens of disk orbit periods ($\tau_{\mathrm{rot}} = 2\pi/\Omega$). The planets gain most of their complement of gas immediately and subsequently sediment heavy element cores and/or sweep up planetesimals over a longer timescale. Disk instability is top down and initially rapid; core accretion is bottom up and initially slow. The critical issue for disk instability is when and where the conditions necessary for fragmentation might actually occur in protoplanetary disks.

We will begin by discussing how GIs operate and what is required for the outcome to be fragmentation. To decide whether real disks fragment, we must rely heavily on detailed numerical simulations.

### 3.1. Gravitational Instabilities: Linear Instability

Gravitational instabilities are caused by the self-gravity of the disk, but, unlike star formation, the onset of planet formation via disk instability is much more strongly influenced by rotation, and it happens in a different thermodynamic environment.

To understand how GIs work, consider first the basic equilibrium of a gas disk orbiting a central star for the case where disk self-gravity cannot be ignored. For the vertical equilibrium in a thin disk where parameters vary smoothly and slowly with orbital distance $a$, equation (3) becomes

$$\frac{1}{\rho}\frac{\partial p}{\partial z} + \left(\frac{GM_\star}{a^2}\right)\left(\frac{z}{a}\right) + 2\pi G\Sigma(a,z) = 0, \quad (49)$$

where $\Sigma(a,z) = \int_{-z}^{z} \rho dz'$ is the surface density of gas at radial position $a$ between $-z$ and $z$, with $z = 0$ being the midplane. In what follows, we use $\Sigma = \Sigma(a)$ to refer to $\Sigma(a,\infty)$. The additional (third) term on the lefthand side of equation (49) characterizes the vertical component of the gravitational field due to the gas disk. The radial equilibrium for a thin disk is then

$$\frac{1}{\rho}\frac{\partial p}{\partial a} - a\Omega^2 + \left(\frac{GM_\star}{a^2}\right) - g_a^D = 0, \quad (50)$$

where $g_a^D$ is the additional radial component of the gravitational field due to disk self-gravity, which depends on the detailed distribution of the disk surface density $\Sigma(a)$. When $H/a$ is small, disk self-gravity can be important in equation (49) without necessarily being a major contribution in equation (50). So, to derive simple relations in later sections, we will sometimes approximate $\Omega$ by using the Keplerian angular speed, $\sqrt{GM_\star/a^3}$.

GIs arise through perturbation of the equilibrium quantified by equations (49) and (50). A classic linear stability analysis by Toomre (1981) for axisymmetric radial ripples is very instructive. Equations (49) and (50) define the zero-order equilibrium state of the disk, which we denote by a subscript zero, e.g., $p_0$, $\rho_0$, etc. Physical variables in the perturbed disk can then be represented by these zero-order quantities plus small perturbations, i.e., $p = p_0 + \delta p$, with $|\delta p| \ll p_0$. To simplify the analysis, the radial wavelength $\lambda$ of the perturbations is taken to be small compared with both $a$ and the scale length of radial changes in $\Sigma$, but large compared with the vertical scale-height $H$. The thin vertical structure of the disk is assumed to remain in hydrostatic equilibrium according to equation (49) even when perturbed. So, after the acceleration term $\partial v_a/\partial t \simeq \partial(\delta v_a)/\partial t$ is introduced into the lefthand side of equation (50) in order to describe the dynamics of the perturbations, the radial equation is integrated over $z$. Equation (50) allows the zero-order terms to be cancelled, and only terms to first order in the perturbed quantities are retained in the linear analysis (for more details, see Binney and Tremaine, 1987).

By looking for wave-like perturbations such that, e.g., $\delta p \propto \exp(i\omega t \pm 2\pi a/\lambda)$, one can derive the dispersion relation

$$\omega^2 = \kappa^2 - \frac{4\pi^2 G\Sigma}{\lambda} + \frac{4\pi^2 c^2}{\lambda^2}, \quad (51)$$

where $\lambda$ is positive-definite and $c^2$ is the sound speed squared. The quantity $\kappa$ in equation (51) is the *epicyclic frequency* at which a perturbed fluid element oscillates about its equilibrium position (Binney and Tremaine, 1987). For a strictly Keplerian disk, $\kappa = \Omega$, because, when the radial equilibrium is dominated by the central force of the star, a slightly perturbed fluid element follows an elliptical orbit with the same period as $\tau_{\rm rot}$. As viewed from its unperturbed equilibrium position, an oscillating fluid element appears to execute a small elliptical "epicycle". The epicycle has the opposite or retrograde sense of rotation compared with the orbital motion of the disk.

Equation (51) reveals a great deal about the behavior of perturbations in self-gravitating disks. As long as the right-hand side (RHS) of equation (51) is positive, $i\omega t \pm 2\pi a/\lambda$ is a complex number and so the ripples are waves which oscillate and propagate radially. However, when the RHS is negative, there are stationary solutions that grow exponentially (the argument of the exponential is real). The disk is then unstable and tends to break apart into dense concentric rings.

An examination of the individual terms on the RHS of equation (51) reveals the stabilizing and destabilizing influences. The first and third terms are always positive and represent stabilization by rotation and gas pressure, respectively. As $\lambda$ gets large, the first term dominates the RHS of equation (51); in other words, rotation stabilizes long wavelengths. Because the central gravitational force of the star appears in equation (51) through $\kappa$, it is also a stabilizing influence. When $\lambda$ is small, the third term dominates; in other words, gas pressure stabilizes small wavelengths. In contrast to the first and third terms, the self-gravity of the disk represented by the second term on the RHS, which is only important at intermediate wavelengths, is always negative and hence is a destabilizing influence. If the acceleration imparted by disk gas, $G\Sigma$, is large enough relative to $\kappa^2 \lambda$ and $c^2/\lambda$, some middle range of wavelengths will be unstable. Because the destabilizing term is due to disk self-gravity, we refer to such instabilities as gravitational instabilities.

Setting $\omega = 0$ in equation (51) and solving for $\lambda$, one can see by simple algebra that there is a range of unstable wavelengths when

$$Q = \frac{c\kappa}{\pi G\Sigma} < 1. \quad (52)$$

Notice that the numerator of the Toomre stability parameter $Q$ is, in effect, a geometric mean of the stabilizing influences of pressure and rotation, while the denominator represents the destabilizing influence of disk self-gravity. Thus, fixing the mass of the star, a protoplanetary gas disk becomes unstable to GIs if it is sufficiently cold (low $c$) or massive (high $\Sigma$). The critical wavelength $\lambda_{\rm cr}$ that first becomes unstable when $Q$ dips below unity is

$$\lambda_{\rm cr} = \frac{2\pi^2 G\Sigma}{\kappa^2} \approx \frac{2\pi H}{Q}, \quad (53)$$

as can be derived by setting $\omega = 0$ and $Q = 1$ in equation (51). For the approximate relation on the right in equation (53), we use $c \approx H\Omega$, which still holds to within a factor of order unity for a self-gravitating disk.

As summarized in various reviews (e.g., Durisen et al., 2007), multi-dimensional simulations of disks show that nonaxisymmetric perturbations, i.e., spiral waves, become unstable for

$$Q = \frac{c\kappa}{\pi G \Sigma} < Q_{\rm cr}, \quad (54)$$

where $Q_{\rm cr}$ is between about 1.5 and 2.0, depending on the detailed structure of the gas disk. Because spiral waves can be unstable in disks that are stable against growth of the axially symmetric ripples we just discussed, spiral disturbances are considered to be more unstable, i.e., a disk which transitions slowly from stable to unstable conditions will first manifest GIs in the form of nonaxisymmetric waves. In fact, in simulations, the linear growth of GIs from noise is characterized by the appearance of multi-armed spirals, which are typically trailing due to the Keplerian shear. These waves grow exponentially from noise in the linear regime on the dynamic timescale $\tau_{\rm rot}$ (e.g., *Nelson et al.*, 1998; *Pickett et al.*, 1998).

With a little algebra, $Q$ in equation (52) can be rewritten approximately as twice the ratio of the star's vertical gravitational field to the disk's vertical gravitational field, namely, $2g_z^\star/g_z^D$, where both $g_z^\star$ and $g_z^D$ are evaluated at $z = H$. So, $Q < 2$ corresponds roughly to $g_z^D > g_z^\star$. As expected, GIs appear when the self-gravity of the gas disk becomes sufficiently important, i.e., when the second and third term on the lefthand side of equation (49) become comparable.

The discussion associated with equation (51) makes it clear how self-gravity could break a disk into ringlets, but how do spiral waves grow? The likely mechanism is explained in *Toomre* (1981) and *Binney and Tremaine* (1987). Consider part of a leading spiral wave spanning its own corotation radius (CR hereafter), the radial location in the disk where the angular pattern speed of the wave equals $\Omega$, i.e., where one arm of the wave goes around the disk in exactly $\tau_{\rm rot}$. Shear in the disk causes the arc to "swing" from leading to trailing. Because $\kappa \approx \Omega$ in a nearly Keplerian disk, the fluid elements of the disk in the crest of the wave near the CR are executing epicyclic oscillations that keep them in the crest. This allows self-gravity to amplify the density maximum at the crest of the wave. Numerical studies of protoplanetary disks in which spiral waves grow show that the conditions are indeed right for the *swing amplification* mechanism to operate (*Pickett et al.*, 1998; *Mayer et al.*, 2004), and the spiral waves that grow always straddle their CR (see. e.g., *Mejía et al.*, 2005).

**3.2. Occurrence of Gravitational Instabilities**

To get a feeling for when GIs are expected in protoplanetary disks, we can evaluate equation (54) for the typical parameters used in earlier sections, namely $H/a \approx 0.05$ at $a = 5.2$ AU for a disk orbiting a solar-mass star (corresponding to a temperature $T \sim 75$ K if $c$ is the adiabatic sound speed). Let us assume that $Q_{\rm cr} \approx 1.7$. Then,

$$Q \approx Q_{\rm cr} \left( \frac{3 \times 10^3 \,{\rm g\,cm^{-2}}}{\Sigma} \right). \quad (55)$$

The typical solids' surface density $\Sigma_s \sim 10\,{\rm g\,cm^{-2}}$ discussed in Section 2 for core accretion models scales to $\Sigma \sim 10^3\,{\rm g\,cm^{-2}}$ for gas with solar composition when half the heavy elements are assumed to be in the solid phase. A surface density $\Sigma = 10^3\,{\rm g\,cm^{-2}}$ at $5.2$ AU is three times the standard Hayashi MMSN (*Hayashi*, 1981), but is only a factor of about 1.5 larger than more recent MMSN models (*Davis*, 2005; *Desch*, 2007). For a disk around a solar-type star to be unstable to gas-phase GIs at Jupiter's orbit radius, i.e., $Q < Q_{\rm cr}$, equation (55) tells us that $\Sigma$ must be $\gtrsim 3 \times 10^3\,{\rm g\,cm^{-2}}$, which is ten times higher than the Hayashi MMSN and four or five times higher than the newer MMSN models.

There are two ways that such a high $\Sigma$ might occur at $5.2$ AU: either the disk is massive everywhere or there is a localized concentration of mass near this radius. In the first case, where $\Sigma(a)$ is relatively smooth, gas-phase GIs at the orbit radii of the gas giants in the solar system require a total protoplanetary gas disk mass of about $0.1\,M_\odot$ or more (*Boss*, 2002a), and disk masses of this order are typical in hydrodynamics simulations used to study GIs in protoplanetary disks. Disks may well have masses this large or larger during the accretion phase, when gas is falling onto the young star/disk system at rates of $10^{-5}\,M_\odot$ per year or higher (e.g., *Vorobyov and Basu*, 2006). Alternatively, a disk might have a localized enhancement of surface density due to accumulation of mass in or at the edges of a *dead zone* (*Gammie*, 1996), where radial transport of mass becomes inefficient. This could lead to episodic eruptive phenomena related to FU Orionis outbursts (e.g., *Armitage et al.*, 2001; *Boley and Durisen*, 2008; *Zhu et al.*, 2009).

Although we have not yet discussed the conditions for fragmentation, it is instructive to estimate how much mass is involved in the GIs. Using equation (53) as an estimate for the radial extent of the instability, we expect the mass in one arm of a $k$-armed unstable wave at $5.2$ AU to be about

$$M_{\rm frag} \approx 2\pi \frac{a\lambda_{\rm cr}\Sigma}{k} \sim \frac{11}{k} M_J. \quad (56)$$

Simulations show that $k$ is typically four or five at the onset of GIs (*Mejía et al.*, 2005; *Boley and Durisen*, 2008) when $M_D \sim 0.14\,M_\star$, and therefore the arms of the spirals would indeed have a gas inventory sufficient to produce a giant planet.

Even if a disk is stable ($Q > Q_{\rm cr}$) at the radius of Jupiter's orbit, it may be susceptible to GIs at larger radii. Suppose that the disk temperature $T \propto a^{-1/2}$ (i.e., $H/a \propto a^{1/4}$, a "flared" disk). We know that $\kappa \approx \Omega \propto a^{-3/2}$, so if $\Sigma \propto a^{-s}$,

$$Q = \frac{c\kappa}{\pi G \Sigma} \propto \frac{T^{1/2}\Omega}{\Sigma} \propto a^{s-7/4}. \quad (57)$$

Thus, for any disk with $s < 7/4$, $Q$ becomes less than $Q_{\rm cr}$ if the disk extends to a sufficiently large radius. For a disk with $s < 2$, $M_D(a) = \int_0^a 2\pi a' \Sigma da' \sim a^{2-s}$, and so the disk mass $M_D(a)$ interior of $a$ diverges as $a$ increases. As a result, if a disk with $s < 7/4$ is extended enough to have GIs in its outer regions, it is likely to have a mass that is a significant fraction of the star's mass. Again, this is most likely to occur during the early accretion phase of the star/disk system. If we put in parameters appropriate for this situation at a few hundred AU around a solar-type star, $M_{\rm frag}$ can become tens of Jupiter's masses or more (*Stamatellos and Whitworth*, 2009).

The above arguments are not very sensitive to the temperature distribution $T(a)$. If we use $T(a) \propto a^0$ or $\propto a^{-1}$ instead, then we get $H/a \propto a^{1/2}$ or $H/a$ equal to a constant, respectively, and $Q \propto a^{s-3/2}$ or $Q \propto a^{s-2}$, respectively. Parameter $Q$ will still become less than $Q_{\rm cr}$ at large enough radii for surface densities that do not fall off too rapidly. During the early accretion phase, one expects that, in spatially extended disks, the gas temperature levels off to a constant value ($T \propto a^0$) at large radii due to envelope irradiation. At small radii, some disks may not be irradiated by their central stars due to shadowing. Such disks would have steeper temperature fall-offs than $a^{-1/2}$ and may not be flared. For more detailed analytic arguments about gravitational instabilities in disks during the accretion phase, see *Clarke* (2009) and *Rafikov* (2009).

### 3.3. Nonlinear Growth of Gravitational Instabilities and Fragmentation

There are two possible nonlinear outcomes for GIs: steady-state balance of heating and cooling or fragmentation into dense clumps. As the spiral waves grow in amplitude and steepen into shocks, they turn some of the ordered rotational energy of the disk into heat. Moreover, gravitational torques due to the predominantly trailing character of the spirals lead to net outward transport of angular momentum and inward transport of mass. This transport causes gravity to do work on fluid elements as they sink into the central gravitational potential and is another important source of heat for the gas. If the disk does not shed this heat, its temperature and hence $c$ inevitably becomes larger, which raises $Q$ and shuts off the instability. Disks do radiate, and so GIs can be sustained in an quasisteady state with an average $Q$ near but somewhat below $Q_{\rm cr}$ where the heating by GIs is balanced by radiative cooling (*Gammie*, 2001; *Lodato and Rice*, 2004; *Mejía et al.*, 2005). *Mejía et al.* refer to this as the *asymptotic* state. GIs grow on a dynamical timescale $\tau_{\rm rot}$ for $Q$ well below $Q_{\rm cr}$, but the growth times become long as $Q \to Q_{\rm cr}$ from the unstable side. So the heating rate due to GIs can adjust to the prevailing radiative cooling rate. In this sense, the thermal physics of the disk controls the limiting amplitude of GIs (*Tomley et al.*, 1991; *Pickett et al.*, 2000).

Let us define a radiative cooling time in a column-wise sense by

$$\tau_{\rm cool} = \frac{\int_{-\infty}^{\infty} E_{\rm int}\, dz}{2\, \sigma_{\rm rad} T_{\rm eff}^4}, \qquad (58)$$

where $E_{\rm int}$ is the gas internal energy density and the effective temperature, $T_{\rm eff}$, characterizes the radiant flux (energy per unit area per unit time) $\sigma_{\rm rad} T_{\rm eff}^4$ out the top of a column through the disk. The factor of two in equation (58) accounts for radiation from both the top and bottom of the disk. This equation makes the approximation that all energy loss from the disk is vertical, which need not be true in complicated dynamical situations.

When $\tau_{\rm cool}$ is comparable to or less than $\tau_{\rm rot}$, the balance of heating and cooling discussed above may not be possible. The gas could cool profoundly between the spiral arms before it is reheated by the shock in the next arm. Another way to say the same thing is that, because the growth time of GIs is $\tau_{\rm rot}$, it is the shortest timescale on which GIs can replenish thermal energy in the gas. There will then be large compressions behind the shocks where the self-gravity of the gas could win out entirely. The arms may then break up into dense clumps, which become bound in the sense that the magnitude of the clump self-gravitational energy exceeds the clump's own internal and rotational energies.

The precise value of $\tau_{\rm cool}$ for which dense clumps form must be determined numerically. The first systematic treatment by *Gammie* (2001) for a simple equation of state in a local patch of a razor thin disk showed that fragmentation occurs for $\Omega \tau_{\rm cool} \lesssim 3$, or equivalently $\tau_{\rm cool}/\tau_{\rm rot} \lesssim 1/2$. Global three-dimensional simulations by several groups using a variety of simulation techniques have verified that the *fragmentation criterion* is

$$\frac{\tau_{\rm cool}}{\tau_{\rm rot}} < f_{\rm frag}, \qquad (59)$$

where $f_{\rm frag}$ depends on the gas equation of state and, if $\tau_{\rm cool}$ is a function of time, on the thermal history of the gas (*Rice et al.*, 2003, 2005; *Mejía et al.*, 2005; *Clarke et al.*, 2007). For ideal gases with ratios of specific heats $\Gamma_1 = 5/3$ and $7/5$, $f_{\rm frag} \approx 1$ and 2, respectively. In other words, a "softer" equation of state in the sense of lower $\Gamma_1$ makes a disk more unstable. This may be relevant, because 5.2 AU is close to where the effective $\Gamma_1$ for molecular Hydrogen is expected to change from $5/3$ at large disk radii to $7/5$ at intermediate radii. When $\tau_{\rm cool}$ is time-dependent and approaches the fragmentation condition slowly, $f_{\rm frag}$ becomes about half as large as when $\tau_{\rm cool}$ is constant, and so the disks become more stable and harder to fragment.

For the special case of gas disks that behave isothermally when perturbed, $\tau_{\rm cool}$ is essentially zero, and, instead of a fragmentation criterion based on the ratio $\tau_{\rm cool}/\tau_{\rm rot}$, one finds that all disks fragment when $Q < Q_{\rm frag}$ where $Q_{\rm frag} \approx 1.4$–$1.5 < Q_{\rm cr}$ (*Boss*, 2000; *Mayer et al.*, 2002; *Johnson and Gammie*, 2003). A classic example of fragmentation in an isothermally evolved disk is shown in Figure 7.

The masses of planets that form in fragmenting simulations are roughly compatible with equation (56), but the process of fragmentation can be violent and chaotic, often leading to multiple interacting protoplanetary clumps. Some clumps merge or shear out, and new clumps can form. No one has yet integrated a large number of fragmenting simulations over long times, but a moderately broad spectrum of final clump masses would probably result in a mass range between tenths to multiples of $M_J$ for disks with $M_D \sim 0.1\,M_\star$ that are a few tens of AU in radius (see, e.g., *Mayer et al.*, 2004). A discussion of fragmentation spectra for disks with more extended radii can be found in *Stamatellos and Whitworth* (2009).

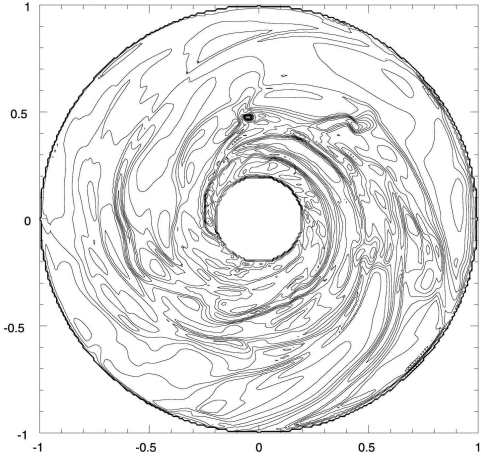

**Figure 7.** Midplane density contours after 374 years for an isothermal simulation of a protoplanetary gas disk spanning 4 to 20 AU around a solar-mass star with $M_D = 0.09\,M_\odot$ and an initial $Q_{\min} = 1.3$. The dense clump near twelve o'clock is gravitationally bound and has about $5\,M_J$. Adapted from *Boss* (2000).

### 3.4. Realistic Radiative Cooling

The pertinent question now is whether $\tau_{\text{cool}}$ for realistic dust opacities is actually short enough to cause fragmentation. The vertical optical depth down into a disk is defined by

$$\tau(z) = \int_z^\infty \bar{\kappa}\rho\,dz', \qquad (60)$$

where $\bar{\kappa}$, with units of cross-section per unit mass, is the mass absorption coefficient for radiation (opacity) averaged in an appropriate way over frequency. If $\bar{\kappa}$ does not vary much with $z$, then the midplane optical depth is $\tau(0) \approx \bar{\kappa}\Sigma/2$. This is essentially the number of photon mean free paths between the midplane and $z = \infty$.

If the radiation emitted by the gas and dust is thermal and if a frequency-independent (grey) opacity characterized by $\bar{\kappa}$ is a good approximation, then momentum conservation for radiation gives (e.g., *Gray*, 1992)

$$\frac{dp_{\text{rad}}}{d\tau} = \left(\frac{a_{\text{rad}}}{4\sigma_{\text{rad}}}\right) F_{\text{rad}} \qquad (61)$$

for plane-parallel geometry, where $p_{\text{rad}}$ is the radiation pressure, $F_{\text{rad}}$ is the net upward flux of radiant energy, and $a_{\text{rad}} \simeq 7.57 \times 10^{-15}\,\text{erg cm}^{-3}\,\text{K}^{-4}$ is the radiation-density constant (notice that the ratio in parentheses in equation (61) is equal to the inverse of the speed of light).

For no radiation shining down on the disk, for a constant upward flux ($F_{\text{rad}} = \sigma_{\text{rad}}T_{\text{eff}}^4$), and for a radiation field that is well described by black-body properties, e.g., $p_{\text{rad}} = a_{\text{rad}}T^4/3$, the standard *grey atmosphere* solution to equation (61) is

$$T^4 = \frac{3}{4}T_{\text{eff}}^4(\tau + q), \qquad (62)$$

where $q$ depends on boundary conditions and approximations but is roughly $2/3$. Equation (62) strictly applies only to a semi-infinite atmosphere with a constant upward flux. A disk is finite and has a flux which varies with $\tau$ (*Hubeny*, 1990). Nevertheless, within a factor of order unity, equation (62) is a useful estimate relating the midplane temperature $T_{\text{mid}} = T(z = 0)$, i.e., $T(\tau)$ at $\tau = \tau(0)$, to the expected effective temperature $T_{\text{eff}}$ in radiative equilibrium given a midplane optical depth $\tau(0)$.

When $\tau(0) \gg q$, the disk is optically thick. In equation (58), one can approximate the integral of $E_{\text{int}}$ to be $3\Sigma\mathcal{R}T_{\text{mid}}/(2\mu)$, where $\mathcal{R}$ is the gas constant and $\mu$ is the mean-molecular weight (see Section 2.5). Then,

$$\tau_{\text{cool}} \sim \frac{9}{32}\left(\frac{\mathcal{R}}{\sigma_{\text{rad}}}\right)\left(\frac{\bar{\kappa}}{\mu}\right)\left(\frac{\Sigma^2}{T_{\text{mid}}^3}\right) \propto a^{(3-4s)/2}, \qquad (63)$$

where, for simplicity, the final dependence on $a$ is based on assuming that $\bar{\kappa}$ is constant and $T_{\text{mid}} \propto a^{-1/2}$. (If $T_{\text{mid}} \propto a^0$ or $\propto a^{-1}$ then $\tau_{\text{cool}} \propto a^{-2s}$ or $\tau_{\text{cool}} \propto a^{3-2s}$, respectively). At 5.2 AU for the usual parameters, equations (60) and (63) give

$$\tau(0) \sim 10^3 \left(\frac{\bar{\kappa}}{1\,\text{cm}^2\,\text{g}^{-1}}\right)\left(\frac{\Sigma}{3 \times 10^3\,\text{g cm}^{-2}}\right) \qquad (64)$$

and

$$\tau_{\text{cool}} \sim 10^5\,\text{yr}\left(\frac{\bar{\kappa}}{1\,\text{cm}^2\,\text{g}^{-1}}\right)\left(\frac{\Sigma}{3 \times 10^3\,\text{g cm}^{-2}}\right)^2. \qquad (65)$$

Note that, for $H/a = 0.05$ at 5.2 AU, $T_{\text{mid}} \approx 75\,\text{K}$ if we use an adiabatic sound speed with $\Gamma_1 = 5/3$ and $\mu \approx 2.4$. In an unstable disk at 5.2 AU, where $\tau_{\text{rot}} \approx 12$ years, equation (65) gives $\tau_{\text{cool}}/\tau_{\text{rot}} \approx 10^4$, which is orders of magnitude greater than $f_{\text{frag}}$ for any reasonable value of $\Gamma_1$.

We have just seen that, for our standard parameters, equations (54), (55), (59), and (63) together imply that, at 5.2 AU, a disk which is dense enough to be gravitationally unstable will cool too slowly to fragment. *Rafikov* (2005) first pointed out this conundrum. Although it depends on detailed assumptions about how disk properties vary with $a$, it can reasonably be argued that this problem extends out to about 40 AU or so, exactly the region where one would like to form the giant planets in our own solar system. Moreover, if we relax our choice of $H/a = 0.05$, then lowering $\tau_{\text{cool}}$ at fixed $\bar{\kappa}$ requires a lower $\Sigma$ and/or a higher $T_{\text{mid}}$, both of which would tend to make the disk more stable.

One can instead imagine lowering $\bar{\kappa}$ itself because dust grows and settles to the midplane, but there is a limit to the efficacy of this ploy. The most efficient cooling by thermal emission occurs for $\tau(0) \approx 1$. For lower $\tau(0)$, the disk is a poor absorber and, hence, by Kirchhoff's law, a poor thermal emitter. All else being equal, equation (64) shows that a reduction of $\tau(0)$ to unity requires a reduction of $\bar{\kappa}$ by a factor of $10^3$. With this reduction, $\tau_{\rm cool}/\tau_{\rm rot}$ becomes about 10, which is still larger than the largest expected value of $f_{\rm frag}$.

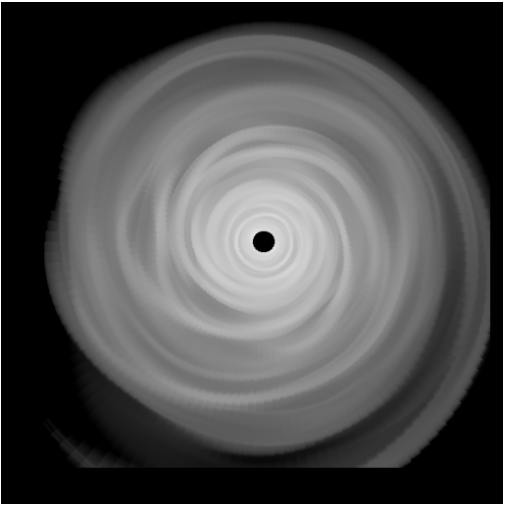

**Figure 8.** Midplane density greyscale of a nonfragmenting and radiatively cooled disk with $M_\star = 0.5\,M_\odot$, $M_D = 0.07\,M_\odot$, and an initial $Q_{\min} = 1.5$ after about 2000 years. The box shown is 80 AU on a side. Adapted from *Boley et al.* (2006).

These analytic arguments suggest that gas giant planet formation by disk instability probably does not work inside several tens of AU. On the other hand, equations (57) and (63) further suggest that the instability criterion (equation (54)) and the fragmentation criterion (equation (59)) may be satisfied simultaneously at large enough radii, roughly 100 AU and beyond, when $0 < s < 7/4$ (see also *Clarke*, 2009; *Rafikov*, 2009).

### 3.5. Computational Results for Disk Instability

#### 3.5.1. Simulations with Realistic Radiative Cooling

The assumptions that go into the results of Section 3.4 are rather simplistic and the details are subject to debate. For instance, $\bar{\kappa}$ is a complicated function of particle size and temperature and $T_{\rm mid}$ at very large orbital radii may reach a constant value. It can also be tricky to define $\tau_{\rm cool}$ properly for the fragmentation criterion in a radiatively cooled disk evolving under the action of GIs (*Johnson and Gammie*, 2003). Disk instability must be tested by detailed multidimensional radiative hydrodynamics simulations. The numerical techniques for such simulations remain difficult, and results to date are disparate and controversial. A detailed discussion of numerical schemes goes beyond the scope of this chapter. Here we only summarize the current situation regarding global simulations that attempt to treat the cooling correctly with realistic opacities. This section deals with simulations of disk instability in regions extending out to tens of AU.

The analytic arguments of Section 3.4 suggest that fragmentation should not occur inside 40 AU or so. Evidence supporting this conclusion comes from simulations using both Smoothed Particle Hydrodynamics (SPH) (*Nelson et al.*, 2000; *Stamatellos and Whitworth*, 2008; *Forgan et al.*, 2009) and grid-based hydrodynamics (*Cai et al.*, 2006; *Boley et al.*, 2006, 2007; *Boley and Durisen*, 2008). These papers employ a variety of radiative algorithms. Figure 8 illustrates a typical midplane density structure for a nonfragmenting simulation. These studies treat 0.5–1 $M_\odot$ stars with a disk mass $M_D \sim 0.1\,M_\star$ and with outer disk radii of ten to tens of AU. The disks do not cool rapidly enough to fragment. In the cases where $\tau_{\rm cool}$ is explicitly calculated, nonfragmentation is consistent with the application of equation (59) for the relevant equation of state. Some efforts have been made by those who do not see fragmentation to evolve similar disks and to test their radiative schemes against analytic expectations, e.g., compare *Boley et al.* (2007) and *Stamatellos and Whitworth* (2008).

Evidence for fragmentation comes from two groups, one using grid-based simulations (*Boss*, 2001, 2002a,b, 2007, 2008) and the other using SPH (*Mayer et al.*, 2007). The disk parameters are similar to those of the nonfragmenting simulations, but these researchers find that their disks fragment into dense, bound clumps, with masses of typically a few $M_J$ (see equation (56)). These groups do not agree in detail on the conditions necessary for fragmentation. *Boss* finds that fragmentation is robust, i.e., fairly independent of many physical and numerical parameters, while *Mayer et al.* find that fragmentation requires somewhat higher disk masses and high mean molecular weight and that it is sensitive to some choices of numerical parameters. *Cai et al.* (2010), using an improved radiative cooling scheme similar to that of *Boley et al.* (2006) but with higher numerical resolution, have now evolved the same initial disk and perturbation as in one of the *Boss* (2007) simulations. No fragmentation occurs. The physical conditions are almost identical, so *Cai et al.* (2010) conclude that the difference really seems to be a matter of radiative algorithms.

The different outcomes are not due to inadequate resolution in the nonfragmenting calculations. For instance, some of the simulations in *Boley and Durisen* (2008) have extremely large grids, achieving resolutions of 0.025 AU over a radial range of about 13 AU. This resolution is more than adequate to detect fragmentation when conditions are artificially forced to satisfy equation (59).

Both *Boss* (2004) and *Mayer et al.* (2007) recognize that purely radiative cooling is not fast enough to satisfy equation (59). Both attribute the fragmentation they see to rapid cooling due to convective transport in the vertical direction and cite upwellings at the sound speed associated with spiral arms in their disks as evidence for thermal convection. However, numerical studies (*Boley et al.*, 2006, 2007) combined with analytic arguments (*Rafikov*, 2007) suggest that the grid-based codes which do not see fragmentation are fully capable of modeling thermal convection and that thermal convection in a disk is unlikely to carry more than tens of percent of the vertical flux, insufficient to reduce cooling times by orders of magnitude. *Boley and Durisen* (2006) explain the upwellings associated with spiral shocks, which are also seen in nonfragmenting simulations, as hydraulic jumps, a dynamic phenomenon very different from thermal convection.

The disagreement about cooling times leads to disagreement about how the disk metallicity affects fragmentation. *Boss* (2002b) studies variations in metallicity by varying $\bar{\kappa}$ up and down by factors of ten relative to solar composition and finds that fragmentation is not sensitive to $\bar{\kappa}$. On the other hand, *Cai et al.* (2006), using simulation techniques similar to those in *Boley et al.* (2006), find that GIs are sensitive to metallicity in simulations where $\bar{\kappa}$ is varied only from 1/4 to 2 times the solar opacity. No simulation in *Cai et al.* produces fragmentation, but the GIs in the lower metallicity disks are stronger, which is expected if $\tau_{\rm cool}$ is mostly determined by radiative processes.

Finally, *Boss* (2002a), *Cai et al.* (2008), and *Stamatellos and Whitworth* (2008) do agree with analytic arguments by *Matzner and Levin* (2005) that sufficiently strong external irradiation of a disk can weaken and suppress GIs. Strongly irradiated disks should not fragment. An easy way to estimate the necessary flux of irradiation $F_{\rm irr} = \sigma_{\rm rad} T_{\rm irr}^4$ to stabilize a disk is to compute the irradiation temperature, $T_{\rm irr}$, such that the resulting sound speed $c(T_{\rm irr})$ satisfies the inequality in equation (54). At 5.2 AU, $T_{\rm irr} \sim 75$ K suffices to suppress GIs altogether for $M_D \approx 0.1\,M_\odot$. No matter what else one might believe about disk instability in the region within tens of AU, it seems pretty clear that irradiation of this order or higher prevents planets from forming by GIs.

### 3.5.2. Enrichment of Heavy Elements

In the standard disk instability model, heavy element cores form *after* the bulk of the gas in the planet becomes a gravitationally bound equilibrium object. Although such protoplanets are likely to be rotating (see, e.g., *Mayer et al.*, 2004), let us approximate them here as hydrostatic equilibrium spheres, with the gas pressure gradient balancing gravity. As discussed in Section 2.1, solid particles respond to gas pressure forces only indirectly through gas drag. Thus, in the absence of turbulence, small solid particles slowly sink to the center of a gas giant protoplanet at a drift speed $v_d \approx g\tau_f$ (see Section 2.1), where $g$ is the magnitude of the gravitational acceleration inside the protoplanet and $\tau_f$ is the friction time. For a newly formed $1\,M_J$ protoplanet with a radius $R \approx 0.5$ AU, approximately half $\lambda_{\rm cr}$ in equation (53) for our standard parameters, $g \approx GM_J/R^2 \simeq 2.3 \times 10^{-3}\,{\rm cm\,s}^{-2}$, and the internal temperature would be $T \sim 100$ K. Using equation (2), one gets a sedimentation time

$$\tau_{\rm sed} \approx \frac{R}{v_d} \sim 6 \times 10^3\,{\rm yr}\left(\frac{1\,{\rm g\,cm^{-3}}}{\rho_{\rm sp}}\right)\left(\frac{1\,{\rm cm}}{R_{\rm sp}}\right), \quad (66)$$

in rough agreement with an estimate of *Boss* (1997). Detailed calculations by *Helled et al.* (2008), starting with micron-sized particles but including aggregation and growth, give similar times. The sedimentation timescale, $\tau_{\rm sed}$, is short compared with the Kelvin-Helmholtz contraction timescale, $\tau_{\rm KH}$, of the protoplanet, which is a few times $10^5$ years (*Helled et al.*, 2006). Turbulence slows sedimentation, but $\tau_{\rm sed}$ for centimeter-sized particles is still shorter than $\tau_{\rm KH}$. Therefore, gas giants formed by disk instability probably are able to produce heavy element cores out of their initial complements of heavy elements. A $1\,M_J$ mass of solar composition could form a $\sim 6\,M_\oplus$ core.

Jupiter and Saturn are strongly enhanced in heavy elements compared with solar composition, as are some transiting exoplanets (e.g., HD 149026 b). There are two ways that gas giants formed by disk instability could end up with complements of heavy elements which are fractionally larger than in their natal gas disk: accreting planetesimals from the surrounding disk after formation or forming in a gas clump already enhanced in solids above the original gas disk composition.

Detailed computations of planetesimal accretion (*Helled et al.*, 2006; *Helled and Schubert*, 2009) for a $1\,M_J$ protoplanet formed by disk instability show that total masses of accreted planetesimals can vary anywhere from 1 to over $100\,M_\oplus$, depending upon the initial location of the planet and upon the assumed mass and radial density distribution of the parental disk. Large accretion rates of heavy elements are made possible by the appreciable starting radii of gas giants in the disk instability model coupled with their moderately long contraction timescale. Such calculations are of course sensitive to assumptions, and a great deal more work is needed, but it appears that heavy element enrichment of planets may not be a strong discriminator between disk instability and core accretion models.

The other intriguing possibility is that, at the time of fragmentation, spiral arms may already be enhanced in solids relative to the ambient disk due to gas drag effects. How might this occur? We have seen that solid particles respond to pressure gradient forces by drifting relative to the gas in a direction opposite to the gradient. For example, particles drift or sediment toward the pressure maximum at the disk midplane (Section 2.1) or at the center of a protoplanet (as discussed above). The same thing applies to radial drift of solid particles in the protoplanetary disk: they drift radially in a direction opposite to the direction of the radial pressure gradient (*Weidenschilling*, 1977), which affect the gas orbital velocity, and concentrate at a radial pressure maximum (*Haghighipour and Boss*, 2003).

The physics in this case is complicated by angular momentum exchange. Equation (50) determines the equilibrium rotation rate of the gas, $\Omega$, which includes the effect of the gas pressure term. Centrifugal balance against gravity in the absence of gas pressure requires a different angular speed for solids, $\Omega_s$, such that $\Omega_s^2 = GM_\star/a^3 - g_a^D/a$. From equation (50) we have that if $\partial p/\partial a < 0$, then $\Omega < \Omega_s$, while if $\partial p/\partial a > 0$, then $\Omega > \Omega_s$. As a consequence, gas drag exerts a torque (per unit mass) on solid particles equal to $(\Omega - \Omega_s)a^2/\tau_f$. Then, by conser-vation of angular momentum, the radial drift velocity is $da/dt = (a/\pi)(\Omega - \Omega_s)(\tau_{\rm rot}/\tau_f)$ (see Section 2.9). In disk regions with a negative radial pressure gradient, an orbiting particle feels a head wind and drifts inward due to loss of orbital angular momentum. In disk regions with a positive radial pressure gradient, a solid particle feels a tail wind and drifts outward due to gain of orbital angular momentum. If there is a pressure maximum at some value of $a$, solids drift toward the pressure maximum from both sides and become concentrated relative to the gas.

The magnitude of the radial drift speed depends on the ratio $\tau_f/\tau_{\rm rot}$. For small particles, $\tau_f/\tau_{\rm rot} \ll 1$ (see equation (4)), and particles are essentially entrained with the gas. The head and tail winds are very small ($\Omega_s \simeq \Omega$), and so the radial drift is slow. Drift speeds increase with particle size in the small particle regime. Very large particles, for which $\tau_f/\tau_{\rm rot} \gg 1$, orbit at angular speed $\Omega_s \neq \Omega$. They can experience considerable head and tail winds, but radial drift speeds are again small because of the long friction times. Drift speeds decrease with particle size in this regime. Maximum radial drift speeds ($\sim a|\Omega - \Omega_s|$) occur for $\tau_f/\tau_{\rm rot} \sim 1$, corresponding to particles of about a meter in radius (*Weidenschilling*, 1977). Such particles can be swept into radial pressure maxima on timescales of only a few hundred years (*Haghighipour and Boss*, 2003).

It is not obvious *a priori* that this phenomenon generalizes to the dynamic environment of a GI-active disk with spiral waves, but, in fact, solid particles of this optimal size do drift into the pressure maxima of spiral arms on similarly short timescales (*Rice et al.*, 2003). This concentration of solids relative to gas in spiral arms can trigger GIs in the solids themselves (*Rice et al.*, 2006) and could even increase the tendency of the gaseous arms to fragment (*Mayer et al.*, 2007). In the latter case, a protoplanet formed by disk instability could initially be enhanced in heavy elements relative to the disk gas.

### 3.6. Bimodal and Hybrid Planet Formation Models

As argued analytically in Section 3.4 and illustrated in Figure 9, even codes that do not produce fragmentation in disks within tens of AU around solar-type stars confirm that fragmentation can and does occur in radially extended and massive disks, beyond about 100 AU (*Boley*, 2009; *Stamatellos and Whitworth*, 2009). Appropriate scaling of equation (56) to large orbital distance and very massive disks leads us to expect the pieces to be super-Jupiters, brown dwarfs, or low-mass stars. This is exciting in light of recent discoveries, discussed elsewhere in this volume (see the chapter by Traub and Oppenheimer), that massive gas giant planets seem to exist at large radii around relatively young stars. Standard core accretion models do not form planets at large orbital distances because it may take too long to form a solid core, whereas violent GIs in an outer disk may occur during the early disk accretion phase for a range of plausible parameters (*Voroboyov and Basu*, 2006; *Clarke*, 2009; *Rafikov*, 2009).

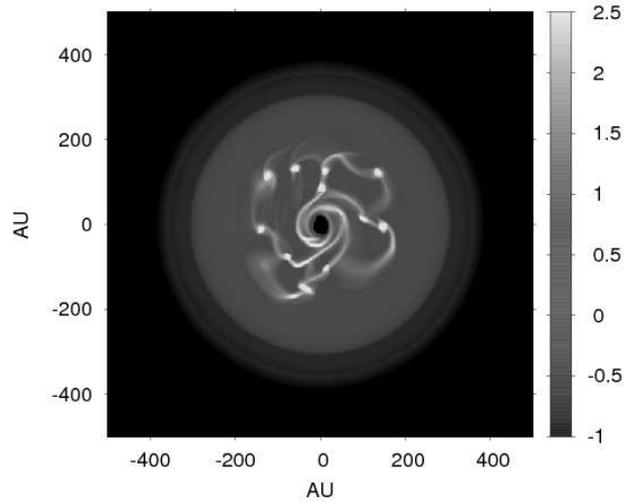

**Figure 9.** Logarithmic surface density (in cgs units) for a radiatively cooled simulation of a fragmenting protoplanetary gas disk around a solar-mass star that is accreting at a rate of $10^{-4}\,M_\odot$ per year. The dense clumps have masses ranging from $4\,M_J$ to $14\,M_J$. Adapted from *Boley* (2009)

Unless planets formed by core accretion can migrate or be scattered to large distances, it seems reasonable to conjecture that planets found in large orbits are formed *in situ* by disk instability. But, if nonfragmentation of GI-active disks in the inner tens of AU prevails as the consensus view, we still have to rely on core accretion to explain the bulk of gas giant exoplanets discovered so far, whose orbits are within a few AU of their stars. This leads to the notion (*Boley*, 2009) that gas giant planet formation may actually be bimodal, with different mechanisms dominating in different regions of protoplanetary disks. Even in this case, it is possible that gas-phase GIs play a role in core accretion by accelerating the formation of solid planetesimals through migration of solids into structures formed in the disk by GIs (e.g., *Haghighipour and Boss*, 2003; *Rice et al.*, 2004, 2006; *Durisen et al.*, 2005).

### 3.7. Summary of Disk Instability Models

Once the necessary conditions exist, namely $Q \lesssim 1.5$–1.7 and $\tau_{\rm cool}/\tau_{\rm rot} < f_{\rm frag} \sim 1$, disk fragmentation into bound clumps of gas is a robust dynamic process that requires only a few to tens of orbital periods. Gas giants formed by disk instability can subsequently produce heavy element cores by sedimentation and can become enriched in heavy elements relative to the background gas disk by accreting planetesimals. Disk instability is expected to occur in the outer regions of extended massive disks during the accretion phase, and it may be the only way to understand the existence of high planetary-mass objects at large distances from their stars, such as Fomalhaut b (*Kalas et al.*, 2008) and HR 8799 b, c, and d (*Marois et al.*, 2008).

Although there is still some controversy, analytic arguments and a number of simulations indicate that disks are unlikely to fragment within a few tens of AU, because the conditions of low $Q$ and a low $\tau_{\rm cool}/\tau_{\rm rot}$ cannot be simultaneously satisfied. The inclusion of stellar irradiation, which tends to stabilize these regions by maintaining a high $Q$, makes the problem worse. Disk instability is most likely to occur in the outer regions of disks relatively early in disk evolution, when disks are massive and still accreting from their protostellar cloud at a high rate. The ultimate fate and survivability of planets formed at such an early phase of disk evolution is unclear. The principle of parsimonious explanations makes appeal to more than one formation mechanism for gas giants unappealing.

## 4. HIGHLIGHTS AND OUTSTANDING QUESTIONS

### 4.1. Solar and Extrasolar Planets

Any theoretical model of giant planet formation has to provide a quantitative explanation for the existence of the four outer planets of the solar system. Although Jupiter's core mass remains uncertain, there is theoretical evidence that Saturn, Uranus, and Neptune have cores of 10–20 $M_\oplus$ (*Podolak et al.*, 1995; *Saumon and Guillot*, 2004). All four planets have massive Hydrogen/Helium envelopes, which represent most of the mass of Jupiter and Saturn and more than 10% of the mass of Uranus and Neptune (e.g., *Guillot*, 2005).

Core accretion models offer a natural explanation for both the formation of heavy element cores and the accretion of massive gaseous envelopes, such as those of Jupiter and Saturn. In this scenario, Uranus and Neptune reached their current masses in an environment that may have been deprived of gases because their cores took too long to form, and never underwent a runaway gas accretion phase. It seems unlikely, instead, that any of these planets formed via GIs in the early solar nebula, principally because theoretical arguments show that fragmentation may only occur beyond many tens of AU from a solar-type star.

Any formation theory also needs to explain the wide range of physical and orbital properties of giant planets around other stars (see the chapter by Cumming). About 6% of Sun-like stars have giant planets orbiting within 4 AU. Jupiter-mass planets in orbit within several AU of Sun-like stars are more common than planets of several Jupiter's masses, and substellar companions more massive than $\sim 10\,M_J$ are rare. Neptune-mass planets may be rather common compared to planets a few times the mass of Neptune. These observational results are broadly predicted by core accretion models, provided that gas in protoplanetary disks survives for at least a few million years. Disk instability models, instead, would predict an abundance of massive objects ($\gtrsim 10\,M_J$) relative to Jupiter-mass planets, since formation begins at very early times in massive disks. GIs, though, provide an appealing prospect for the formation of $\sim 10\,M_J$ objects observed at distances $\gtrsim 100$ AU from their stars, where core formation timescales are probably too long.

Exoplanet data show a strong correlation between stellar metallicity and frequency of giant planet detections (e.g., *Udry and Santos*, 2007): $\sim 25\%$ of stars whose metallicity is twice that of the Sun host a gas giant (within a few AU). This percentage reduces to only $\sim 5\%$ for solar metallicity stars. Disk instability models do not seem to predict this trend. In fact, there is some evidence that GIs are stronger in lower metallicity disks (*Cai et al.*, 2006). Core accretion models do predict this trend (*Kornet et al.*, 2005; *Mordasini et al.*, 2009), provided that the density of solids in a disk is proportional to the metallicity of the central star.

Planet searches around M dwarf stars ($M_\star \lesssim 0.5\,M_\odot$) seem to suggest that the occurrence of giant planets is less likely than it is around more massive stars (*Johnson et al.*, 2007; *Eggenberger and Udry*, 2010). Although these surveys are still incomplete, there is mounting observational evidence that planets of mass $M \lesssim 0.1\,M_J$ orbiting M dwarfs may be common (*Forveille et al.*, 2009). If confirmed by more complete statistics, a correlation between stellar masses and the masses of hosted planets may provide valuable information on the process of planet formation. According to core accretion models, gas giants orbiting low-mass stars should be rare (*Kennedy and Kenyon*, 2008), whereas Neptune-mass planets should be common (*Laughlin et al.*, 2004). According to disk instability models, if conditions are appropriate for disk fragmentation, the occurrence rate of giant planets should not depend strongly on the stellar mass (*Boss*, 2006).

Interpretations of observational data based on either of these mechanisms remain difficult and predictions are still rather qualitative. There is a number of unresolved issues that prevent us from making more quantitative predictions. Such issues are generally related to poor constraints on the initial conditions used in the formation models. Some of these open questions are outlined below.

What factors determined the final masses of Jupiter and other gas giants? The mass of Saturn may have been influenced by the presence of Jupiter and similar conditions may have operated in some extrasolar planetary systems with multiple gas giants. However, a more general answer to this question will require a much deeper understanding of the physical conditions that existed in the solar nebula and other protoplanetary disks.

Are there peaks in the mass distribution function of massive planets? During the final phases of growth, giant planets are in a symbiotic relationship with their parent disks. Therefore, we need better constraints on masses, temperatures, and lifetimes of protoplanetary disks to make predictions on the relative abundances of giant planet masses.

What is the occurrence rate of gas giants and Neptune-mass planets as a function of the stellar mass? What is the percentage of red dwarfs that host massive planets? Currently, a handful of cases are known, as the two Jupiter-mass planets orbiting Gliese 876 (http://exoplanet.eu/), but conclusions cannot be drawn from such small-number statistics.

Is there a numerous population of gas giants orbiting at large separations from their host stars, as GIs models seem to predict? If so, how do their orbital properties and envelope composition differ from those of other known gas giants revolving on smaller orbits? What is the likelihood that they formed closer to the star (perhaps via core accretion) and were later scattered or migrated to larger distances?

### 4.2. Impact of Orbital Migration

Planetary cores of $\sim 10\,M_\oplus$ that undergo type I migration may be subject to rapid orbital decay. One of the main advantages of disk instability models is that they would bypass this phase of planetary growth altogether and avoid this potential threat. Planets formed via GIs, instead, may be subject to prolonged periods of type II migration. But, how relevant is planetary migration during planet formation?

Although a complete answer is not presently known, there is evidence that substantial migration may occur during the formation process. A significant fraction of multi-planet extrasolar systems are close to or in a mean-motion resonance (*Udry and Santos*, 2007). Since it is unlikely that such orbital configurations can be produced by *in situ* formation models, these systems may lend support to convergent migration hypotheses. Recent studies aimed at explaining the structure of the outer solar system (e.g., *Morbidelli et al.*, 2009) suggest that Jupiter and Saturn may have undergone convergent migration, as a result of which Saturn became temporarily captured into a mean motion resonance with Jupiter.

The presence of Neptune-mass planets and gas giants orbiting stars within a few tenths of AU may also constitute evidence of orbital migration, although alternative interpretations are possible. Planet-planet scattering and Kozai cycles (see the chapter by *Lubow and Ida*) can, in principle, drive gas giants very close to their host stars. However, while these last two mechanisms could produce a large misalignment between the stellar spin axis and the planet's orbit axis, disk-induced orbital migration would preserve spin-orbit alignment. Some of the transiting planets allow for the measurement of the sky-projected angle between stellar spins and orbit axes (see the chapter by *Winn*). Currently, about two-thirds of these measurements are consistent with a close spin-orbit alignment (*Simpson et al.*, 2010), suggesting that migration by tidal interaction with the parent disk drove these planets close to their stars.

The standard theory of orbital migration by disk-planet interactions seems able to reproduce the main features of the orbital period distribution of exoplanets (e.g., *Armitage*, 2007), which suggest that considerable migration may have occurred. Yet, the question of whether or not substantial migration is the norm can only be assessed once observations of long-period planets become available. A more complete distribution function of orbital periods can also provide quantitative indications on whether current estimates of the migration rates of Neptune-mass planets are too rapid and, if so, to what extent they are.

## 5. FUTURE PROSPECTS

To achieve a more comprehensive understanding of the formation of planetary systems, at least some of the questions highlighted in Section 4 need be addressed. We envision that answers to those questions can indeed be found in the near future, as further progress is made along three main avenues: data collection from the solar system; observations of star-forming regions, protoplanetary disks, and exoplanets; theoretical studies of star and planet formation.

Interplanetary spacecrafts (e.g., *Voyager*, *Galileo*, and *Cassini*) have provided a wealth of information about the origins and evolution of the outer solar system. The *Juno* mission may finally reveal the mass of Jupiter's core by direct measurements of the planet's gravity field. Although it seems implausible that such information alone will provide definitive evidence for one or the other formation scenario, it is essential for refining interior models of the gas giant nearest to us.

To better constrain current estimates of sizes, masses, dust contents, and lifetimes of protoplanetary disks, it is important to know the detailed physical conditions that exist at various stages of a disk's evolution, starting from the latter phases of stellar formation. The observation of star-forming regions and disks surrounding young stars plays a pivotal role in establishing more comprehensive and accurate models of disk evolution. These models can be then applied in the context of increasingly sophisticated calculations of planet formation that include disk dispersal.

Space observatories, like *Hubble* and *Spitzer* space telescopes, have provided valuable data that has significantly improved our knowledge of star-forming regions and planet-forming environments. Over the next decade, a lot more is expected to be learned about protoplanetary disks with both existing and new-generation instruments, such as the *Atacama Large Millimeter Array* (*Tarenghi*, 2008), the *Stratospheric Observatory for Infrared Astronomy* airborne observatory (*Gehrz et al.*, 2009), and the *James Webb Space Telescope* (http://www.jwst.nasa.gov/). These facilities, among others, will improve in completeness the mass distribution function of Neptune-mass and larger planets. They will likely improve our knowledge on the relationships between stellar metallicity and dust content of young disks. They should also allow direct imaging of high planetary-mass objects and low-mass brown dwarfs, orbiting at several tens of AU from stars, and provide precious data on the abundances of heavy elements and on the chemical compositions of their envelopes.

Ground-based interferometers (e.g., *Very Large Telescope* and *Keck* telescope) and astrometry space observatories, such as *Gaia* (http://sci.esa.int/GAIA) and the *Space Interferometry Mission* (http://sim.jpl.nasa.gov/), will extend the orbital period distribution function of detected exoplanets towards longer-period orbits. Correlations between planet detection frequency and stellar metallicity and between stellar mass and planet mass are also expected to become more accu-rate and statistically significant. If a significant fraction of M dwarf stars possess gas giants, observational data should allow us to compare their orbital and physical properties with those of planets orbiting more massive stars. If a large population of high-mass planets is found at large separations, their atmospheric composition can be compared to those of transiting planets, whose number should also grow significantly thanks to space missions such as *COROT* (http://smsc.cnes.fr/COROT/), *Kepler* (http://kepler.nasa.gov/), and *TESS* (http://space.mit.edu/TESS). Within the next five to ten years, we should be able to test the bimodal distribution hypothesis for gas giant formation based on statistical analysis of orbital properties, mass distribution functions, and envelope compositions.

The trend of exoplanet detections and characterization witnessed over the past decade will carry forward to the next decade. Theoretical studies need to tackle still unsolved problems on the timescales and physical conditions required for gas giant formation. Core accretion models must be extended to include proper feedback on the protoplanetary disk and tidally-induced orbital migration in a self-consistent manner. Disk instability models need to be refined in order to make quantitative predictions on survivability, final orbital radii, and final masses of planetary-mass objects formed after disk fragmentation. As data of better and better quality become available, it may be challenging, from a theoretical standpoint, to combine all pieces in a coherent framework. Some of the pieces now regarded as important may be discarded and replaced by others. Yet, the prospects are in place to turn a blurred picture into one with much higher definition within the next ten years.

**Acknowledgments.** We thank Aaron C. Boley, Uma Gorti, and two anonymous reviewers for useful comments. G.D. acknowledges support from NASA Origins of Solar Systems (OSS) Program grants NNX08AH82G and NNX07AI72G. J.J.L. was supported by NASA OSS grant 811073.02.07.02.4. R.H.D. was partially supported during manuscript preparation by NASA OSS grants NNG05GN11G and NNX08AK36G. G.D. acknowledges computational resources provided by the NASA High-End Computing (HEC) Program through the NASA Advanced Supercomputing (NAS) Division at Ames Research Center.